\documentclass[prx,amsmath,twocolumn,showpacs]{revtex4-1}
\usepackage{epsfig}
\usepackage{epsfig}
\usepackage{graphicx}
\usepackage{epsfig}
\usepackage{rotating}
\usepackage{amssymb}
\usepackage{amsmath}
\usepackage{graphicx}
\usepackage{amsmath}

\usepackage[caption=false]{subfig} 
\usepackage[linesnumbered,ruled]{algorithm2e}
\usepackage{algorithmic}

\usepackage{amssymb}
\usepackage{amsmath}
\usepackage{color}
\usepackage{units}
\usepackage{enumitem}
\def\beq{\begin{eqnarray}}
\def\eeq{\end{eqnarray}}
\newcommand{\be}{\begin{equation}}
\newcommand{\ee}{\end{equation}}
\newcommand{\bea}{\begin{eqnarray}}
\newcommand{\eea}{\end{eqnarray}}

\begin{document}

\preprint{APS/prx}
\title{Revisiting the challenges of max-clique}
%
%
\author{Raffaele Marino}
\email{marino.raffaele@mail.huji.ac.il}
\author{Scott Kirkpatrick }
\email[Corresponding author : ]{kirk@cs.huji.ac.il}
\affiliation{The Hebrew University of Jerusalem, School of Computer Science and Engineering, Edmond Safra Campus, Givat Ram, Jerusalem 91904, Israel}

%

%

%
%
\begin{abstract}
The MaxClique problem, finding the largest complete subgraph in an Erd{\"o}s-R{\'e}nyi $G(N,p)$ random graph in the large $N$ limit, is a well-known example of a simple problem for which finding any approximate solution within a factor of $2$ of the known, probabilistically determined limit, appears to require P$=$NP.  This type of search has practical importance in very large graphs.  Algorithmic approaches run into phase boundaries long before they reach the size of the largest likely solutions.  And, most intriguing, there is an extensive literature of \textit{challenges} posed for concrete methods of finding maximum naturally occurring as well as artificially hidden cliques, with computational costs that are at most polynomial in the size of the problem.

We use the probabilistic approach in a novel way to provide a more insightful test of constructive algorithms for this problem.  We show that extensions of  existing methods of greedy local search will be able to meet the \textit{challenges} for practical problems  of size $N$ as large as $10^{10}$ and perhaps more.  Experiments with  spectral methods that treat a single large clique of size $\alpha N^{1/2}$ \textit{planted} in the graph as an impurity level in a tight binding energy band show that such a clique can be detected when  $\alpha \geq \approx1.0$.  Belief propagation using a recent \textit{approximate message passing} (\textbf{AMP}) scheme of inference pushes this limit down to $\alpha \sim \sqrt{1/e}$.  Exhaustive local search (with early stopping when the planted clique is found) does even better on problems of practical size, and proves to be the fastest solution method for this problem.

\end{abstract}

\maketitle              

\section{Introduction}
\label{sec:introduction}

Phase transitions in the asymptotic behavior of combinatoric problems on random ensembles once were but are no longer surprising. 
Large scale data structures, such as graphs, arise in practical examples. Effective tools for managing them have commercial value. Unlike phase transitions in the physics of materials, the model system and the interactions which couple its elements are known or can be defined. While exact methods can solve only very small examples, simulation of medium scale problems is accessible and may reach very large scale. Methods such as 
finite-size scaling analysis expose regularities \cite{kirkpatrick1985statistical}.  Classic examples include the Satisfiability problem in its many variants \cite{selman1993local,kirkpatrick1994critical}. In this paper, we consider finding maximum cliques in random graphs, specifically Erd{\"o}s-R{\'e}nyi \cite{erdos1959random,bollobas1998book}  graphs of the $G(N,p)$ class, with $N$ nodes (or sites) and each edge (or bond) present with probability $p$.   We further specialize to the case $p = 1/2$, which has the advantage that since the maximum clique is also the maximum independent set (IS) on the complement of the graph, finding the maximum clique at $p = 1/2$ also solves a second famous problem, finding a maximal independent set, since that is the maximal clique on the complement of a graph and the set $G(N,p)$is its own complement.

\begin{figure}
\centering
\includegraphics[width=1\columnwidth, keepaspectratio=true, angle=0]{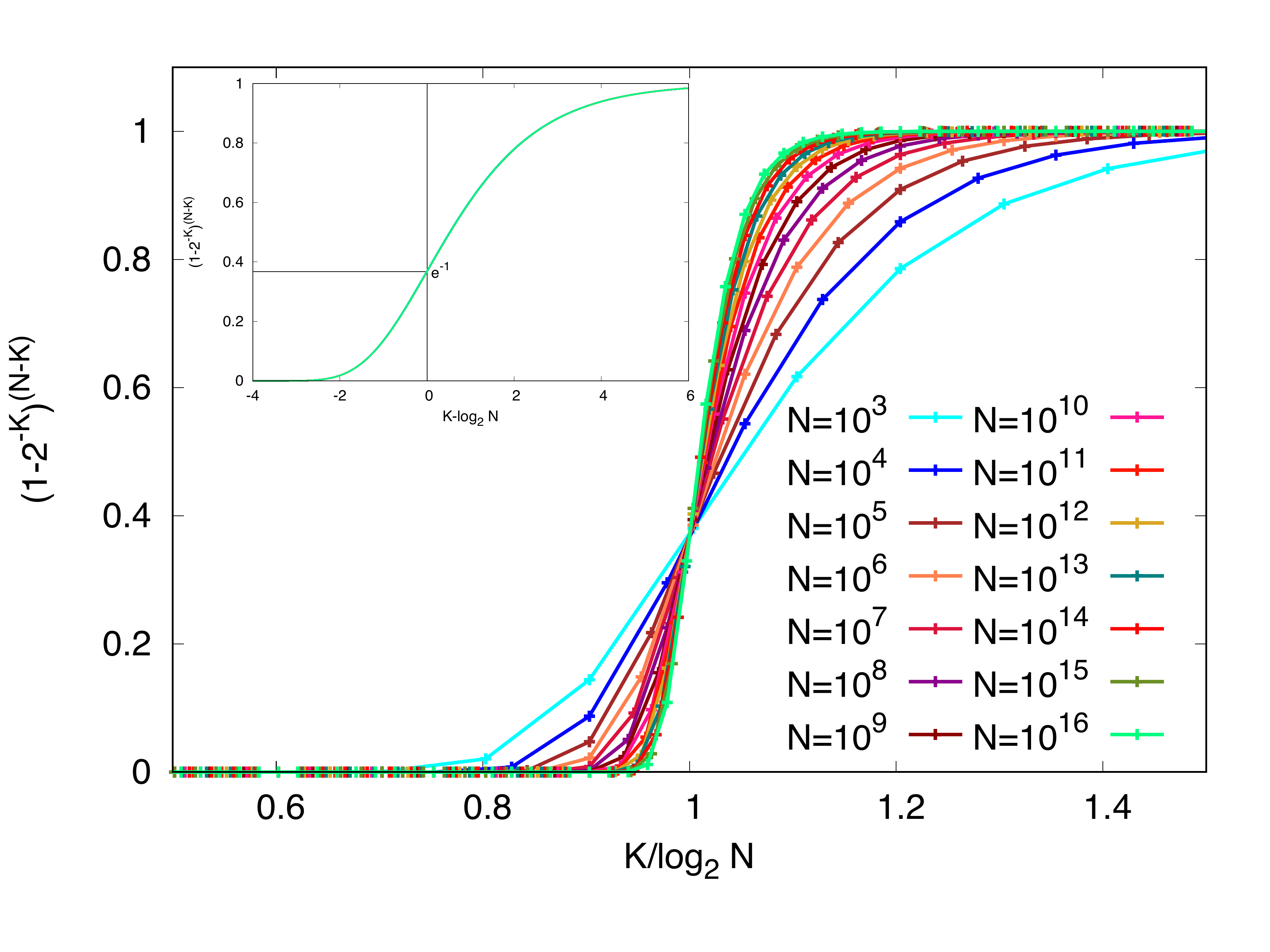}
\caption{The picture shows the probability that we can find no other site to grow the clique to size $K+1$ as function of $K/\log_2 N$. 
The inset shows the universal form that all of these curves take when rescaled.}
\label{figexpanding}
\end{figure}

We will see below that MaxClique is an unusually difficult problem. Naive solution methods can construct cliques of size $\log_2 N$, yet probabilistic arguments show that solutions asymptotically of size $2\log_2 N$ must exist.  No polynomial algorithms that will construct true maximum cliques for arbitrarily large values of $N$ are known. The failure is general, not merely a problem for the rare worst case. This difficulty has been formalized as several challenges, which are difficult to resist.  We shall test several algorithms at large finite values of $N$, to see if the \textit{impossible} can be achieved or approached over a useful range of $N$.  

Greedy methods are fast, but naive and a good starting point for our discussion. 
Start with a site anywhere in the graph, and discard the roughly half of the sites that are not neighbors. 
Pick a neighbor from the \textit{frontier} of the first site that remains.  Then discard the half of the remaining sites that are not a neighbor of the new site.  
Continue in this way until the \textit{frontier} vanishes -- no candidates to extend the clique remain.  
Since we have halved the size of the frontier at each step, it is unlikely that this process can proceed beyond $\log_2 N$ steps.

Let's look at this more precisely and as a function of the scale, $N$, the order of the graph. 
Assuming we have a clique of size $K$, the probability that we can find no other site to grow the clique to size $K+1$ is $(1-2^{-K})^{(N-K)}$, 
as shown in Fig.\ref{figexpanding}, where use of a common scale $K/\log_2 N$, brings the various curves all together at a probability of $\text{e}^{-1}$ when $K$ is equal to $\log_2 N$.  All cliques are extendable when this ratio goes to zero, and we shall see shortly that none are when it exceeds $2$. The slopes of these curves are each proportional to $\log_2 N$.  The simple expedient of plotting the curve for each value of $N$ against $K - \log_2 N$ collapses all of them to a universal limiting form, which is shown in the inset to Fig. \ref{figexpanding}. This is finite-size scaling just as described in \cite{kirkpatrick1985statistical}.  This sort of limit to an algorithm's effectiveness has been called a \textit{dynamic} phase boundary in the literature \cite{krzakala2007gibbs}.  It also shows that cliques constructed by this naive greedy algorithm start to be non-extendable at a size two sites below the dynamic phase boundary and are almost never extendable four sites above, with a functional form that is almost independent of $N$.  Because this threshold occurs at each  $N$ for which $K = \log_2 N$ is an integer, and has a width independent of $N$, it is less "\textit{sharp}" than the phase transitions seen in models of magnetic ordering.

\subsection{History}

Matula first called attention to several interesting aspects of the MaxClique problem on $G(N,p)$.  From the expected number of cliques. $\mathbb{E}(K)$, of size $K$ at $p = 1/2$ \cite{matula1970complete}:

\begin{equation}
\label{eq::averageclique}
\mathbb{E}(K) = {N \choose K} 2^{-{K\choose2}} ,
\end{equation} 

using Stirling's approximation, one can see that this is large at $K = \log_2 N$ but becomes vanishingly small for $K > 2\log_2 N$, providing an upper bound to $ K$.  Matula identified $K_{\text{max}}$ as the largest integer such that 
\begin{equation}
\label{eq::KMAX}
\mathbb{E}(K_{\text{max}} ) \geq 1 .  
\end{equation}
and \cite{matula1970complete,matula1972employee}  expanded the finite $N$ corrections to the continuous function $R(N)$ which solves $\mathbb{E}(R) = 1$ :

\begin{equation}
R(N)=2\log_2 N-2\log_2 \log_2 N+2 \log_2 e/2 +1.
\label{MatBoleqlong}
\end{equation}

\begin{figure}
\centering
\includegraphics[width=1\columnwidth, keepaspectratio=true, angle=0]{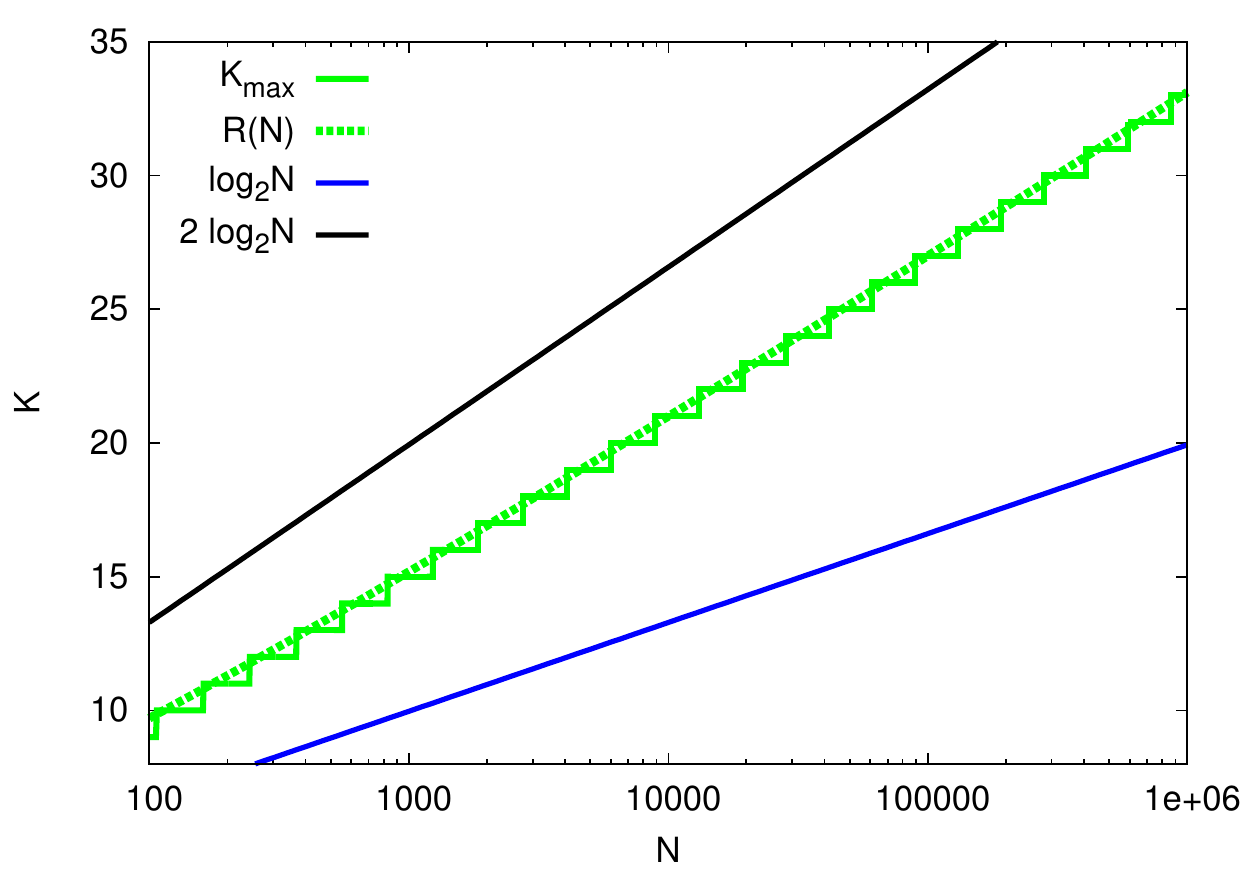}
\caption{Maximum size of a clique in G(N,0.5) as function of the order of the graph. In green are presented $K_{\text{max}}$, result of  solving (\ref{eq::KMAX}) and $R(N)$ given by  (\ref{MatBoleqlong}), respectively. In black is plotted the value of $2 \log_2 N$ and in blue, $\log _2 N$. }
\label{Matstaircase}
\end{figure}

This formula is also discussed in Bollob{\'a}s and Erd{\"o}s \cite{bollobas1976cliques} and by Grimmett and McDiarmid \cite{grimmett1975colouring}. Very tight limits are known showing that $R(N)$ differs by less than $1$ from  $K_{\text{max}}(N)$ as $N \to \infty$.  We will focus on $K_{\text{max}}$, the predicted actual maximum clique size.  In effect, its value follows a staircase with prediction (\ref{MatBoleqlong}) passing through the risers between steps, as shown in Fig. \ref{Matstaircase}.

In his first pa.pdfper, Matula drew attention to what is now termed a \textit{concentration} 
result for the clique problem.  As $N \to \infty$ the sizes of the largest cliques that will occur are concentrated on just two values of $K$, the integers immediately below and above $R(N)$.  To do this, he used the second moment of the distribution of the numbers of cliques of size $K$ to bound the fraction of graphs with no such cliques, and sharpened the result \cite{matula1976largest} by computing a weighted second moment.    In effect, Markov's inequality provides upper bounds, and Chebyscheff's inequality provides lower bounds on the existence of such cliques.  In principle, more detailed evaluations of higher moments could characterize the frequency with which cliques of size $K_{\text{max}}$ are found, but we shall use only Matula's results for the two values of $K$ on which the maximum cliques concentrate. The probability that the maximum clique size is $K_{\text{max}}$ was given by Matula  \cite{matula1970complete,matula1972employee,matula1976largest}.  The fraction of graphs $G(N,p)$ with  maximum  clique size $K_{\text{max}}$, is bounded as follows:
\begin{equation}
\label{eq::2}
\resizebox{0.43\textwidth}{!}{$\left(\sum_{j=\text{max}\{0, 2k-N\}}^k \frac{{N-k \choose k-j}{k \choose j}}{{N \choose k}} p^{-{j \choose 2}}  \right)^{-1}\leq \text{Prob}(K_{\text{max}}\geq k) \leq {N \choose k} p^{k \choose 2}.$}
\end{equation}

\begin{figure}
\centering
\includegraphics[width=1\columnwidth, keepaspectratio=true, angle=0]{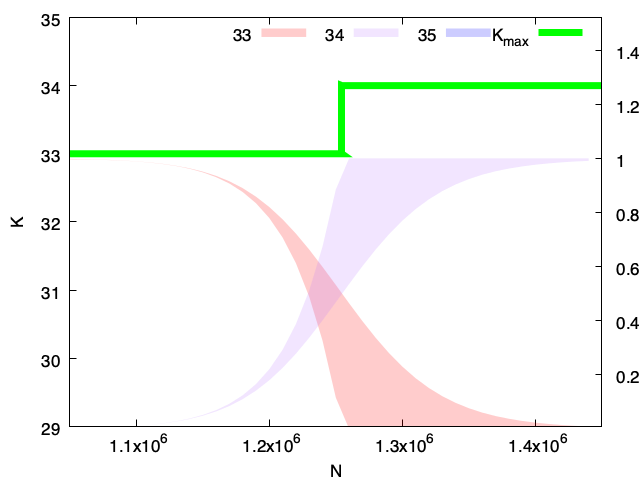}
\caption{The picture shows regions defined by upper and lower bounds (obtained by equation (\ref{eq::2})), on the fractions of graphs of size $N$ with maximum cliques of size $K_{\text{max}}$ and $K_{\text{max}} + 1$. In the large N limit, at a step, the probability for finding each of the two possible sizes becomes $1/2$.}
\label{bigMatprob}
\end{figure}

This leads to the following picture, evaluated for  
large $N$, e.g. $N \approx 10^6$ in Fig. \ref{bigMatprob}, 
we see that at the step between two integer values of $K_{\text{max}}$, half of the graphs will have a few cliques of the new larger value from the upper step, and half will have only cliques with the smaller value from the lower step, but many of them.

\subsection{Challenges}

The traditional approach to surveying and challenging the developers of algorithms for solving hard problems is to assemble a portfolio of such problems, some with a known solution, and some as yet unsolved.  The DIMACS program at Rutgers carried out such a challenge in the mid $1990$'s \cite{johnson1996cliques}.  Roughly a dozen groups participated over a period of a year or more, and the sample graphs continue to be studied.  The largest graphs in the portfolio were random graphs of size $1000$ to $4000$, and the methods available gave results for the largest of these which fell at least one or two short of $R(N)$. (A few did much worse.)  As a result, the actual values of $K_{\text{max}}$ for many test graphs are still unknown.  We argue that a better test for these algorithms on random graphs is to determine to what extent they can reproduce the predicted distribution of results that we see in Fig. \ref{bigMatprob}, both the steps in $K_{\text{max}}$ and the fraction of graphs with each of the dominant values of $K_{\text{max}}$ as it evolves with increasing $N$.  

Several authors have proposed that the search for powerful, effective clique-finding algorithms could be expressed as a \textit{challenge}, perhaps to attract the widest set of challengers to the problem.  
Mark Jerrum, in his $1992$ paper \textit{"Large Cliques Elude the Metropolis Process"}, \cite{jerrum1992large} sets out several of these. 
His paper shows that a restricted version of stochastic search is unlikely to reach a maximum clique, and also introduces the additional problem of finding an artificially \textit{hidden clique}, which we discuss in a later section.  A hidden clique or \textit{planted solution}, is just what it sounds like, a single subgraph of $K_{HC}$ sites, with $K_{HC} > K_{\text{max}}$, so that it can be distinguished, for which all the missing bonds among those sites have been restored. 
A series of papers \cite{alon1998finding,dekel2014finding} show that if $K_{HC}$ is of order $N^{\alpha}$ with $\alpha > 0.5$, a small improvement over our naive greedy algorithm ($SM^{0}$ introduced in the next section) will find such a hidden clique.  

Jerrum's first challenge is to find a hidden subgraph of size $\sim N^{0.5 - \epsilon}$ with probability $> 1/2$, using an algorithm whose cost is polynomial in the number of bonds in the graph (i.e. $N^2$ is considered to be a linear cost).  Jerrum's paper and several others have also turned the identification of any naturally occurring clique larger than the dynamic threshold size into such a challenge: find any clique of size exceeding $\log_2 N$ with probability exceeding $1/2$.  We saw in the discussion of Fig.\ref{figexpanding} that finding cliques which exceed $\log_2 N$ by a small constant number of sites should be straightforward at any value of $N$.  We shall see that both challenges are in fact easy for large, finite and thus interesting values of $N$, and will attempt to characterize for what range of $N$ they remain feasible.

\section{Greedy Algorithms}
\label{sec:maximumcliqueproblem}

In this section we describe the performance of a family of increasingly powerful greedy algorithms for constructing a maximal clique on an undirected graph.  Those algorithms are polynomial in time and use some randomness, but they are myopic in generating optimal solutions. However, because they are relatively fast, significant research efforts has been devoted to improving their performance while adding minimal complexity. We will show ways of combining several of these simple greedy algorithms, to obtain better solutions at somewhat lower cost. 

We start by considering a simple family of greedy algorithms, designated by Brockington and Culberson \cite{brockington1996camouflaging}, as $SM^{i}$, $i=0,1,2..$. $SM^{0}$ improves over the naive approach we described at the outset \cite{kuvcera1991generalized}, by selecting at each stage the site with the largest number of neighbors to add to the growing clique. If there are many such sites to choose from, each connected to all of the sites in the part of the clique identified to that point, one is chosen at random, so multiple applications of $SM^{0}$ will provide a distribution of answers for a given graph $G$. At each stage this choice of the site to add retains somewhat more than half of the remainder of the graph, $Z$, so the resulting clique will be larger than $\log_2 N$, for all finite $N$.  $SM^0$ can be implemented to run in $\mathcal{O}(N^2)$ time.

\begin{figure}
\includegraphics[width=1\columnwidth, keepaspectratio=true, angle=0]{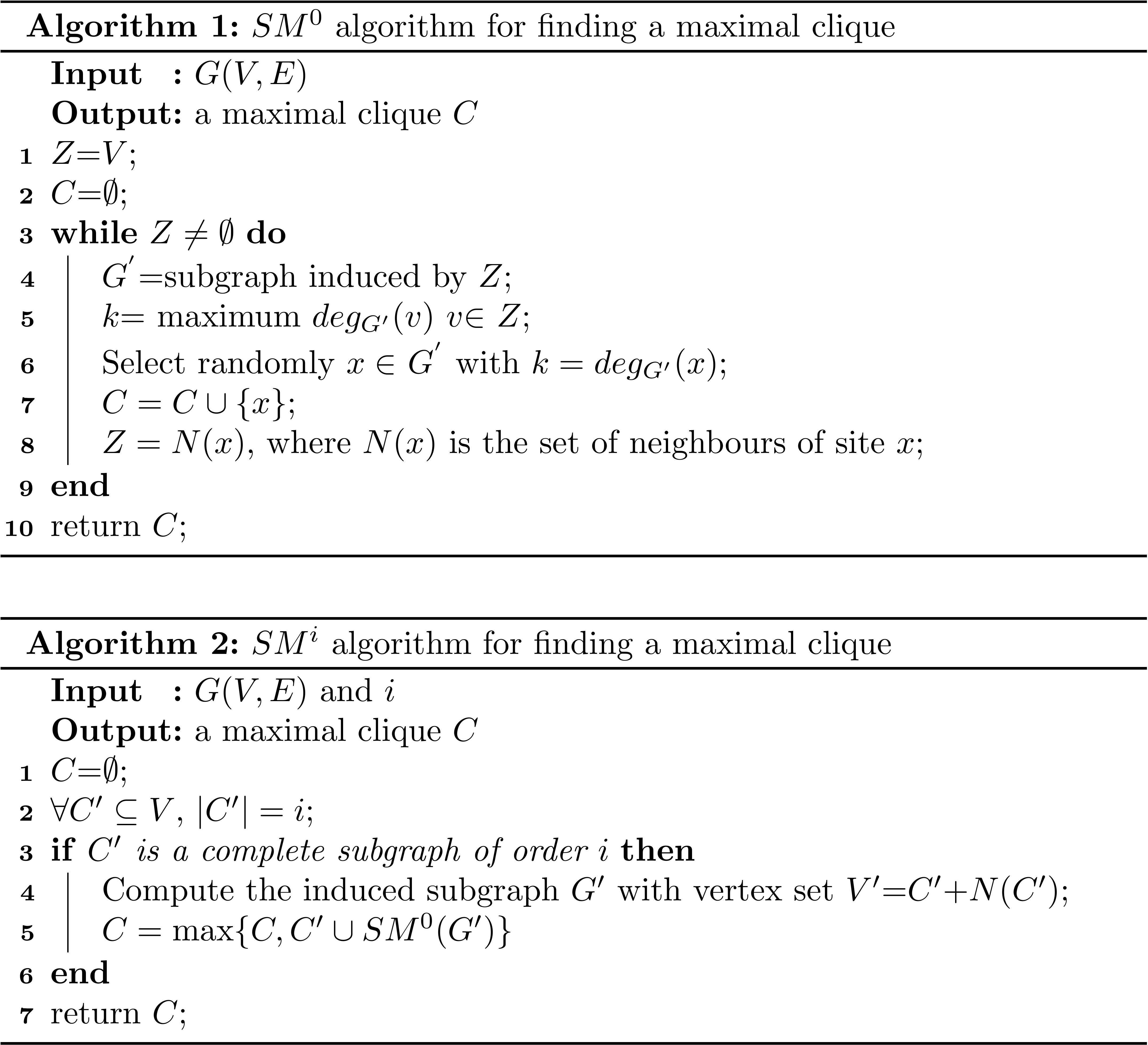}
\label{algo1-2}
\end{figure}


    


$SM^{i}$ for $i=1, 2,...$ are algorithms in which we start our greedy construction with each combination of $i$ vertices which form a complete subgraph, then extend them one site at a time using $SM^0$. In other words, $SM^{0}$ is run starting with each of ${N \choose i} p^{{i \choose 2}}$ complete subgraphs of order $i$. $SM^{1}$, starting with every site, can be implemented to run in $\mathcal{O}(N^3)$. The complexity of $SM^{2}$, which uses all connected pairs, is $\mathcal{O}(N^4)$. The computational complexity for the class of algorithms $SM^{i}$ is $\mathcal{O}(N^{i+2})$.

\begin{figure}
\centering
\includegraphics[width=1\columnwidth, keepaspectratio=true, angle=0]{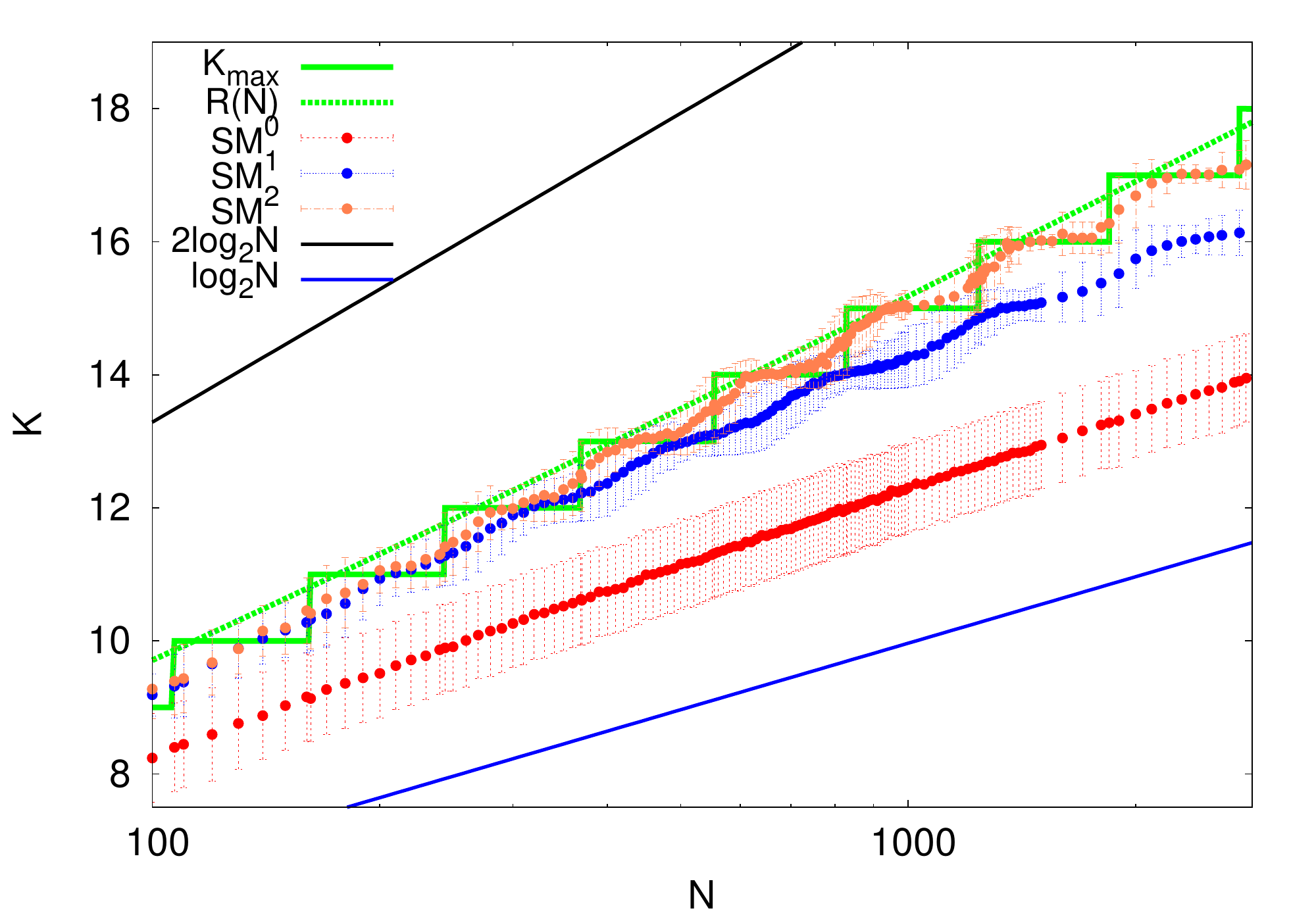}
\caption{Mean and standard deviations of sizes of the maximal cliques from E-R graphs $G(N, p=0.5)$, using the algorithms $SM^{i}$, with $i=0, 1, 2$. The maximum size of a clique as function of the order of the graph is represented by the green staircase $K_{\text{max}}$. The dashed green straight line is $R(N)$ from equation (\ref{MatBoleqlong}). The black and the blue straight lines show $2\log_2 N$ and $\log_2 N$, respectively. Red data points describe the mean maximal clique size obtained by $SM^{0}$, averaging over $2000$ E-R graphs. Blue points describe the maximal clique size found with $SM^{1}$, while orange points are the results of $SM^{2}$. The average is obtained over $500$ and $100$ graphs for $SM^{1}$ and $SM^{2}$, respectively. }
\label{figSM0SM1SM2}
\end{figure}

In Fig. \ref{figSM0SM1SM2} we show the sizes of the maximal cliques on E-R graphs $G(N, p=0.5)$, found using the algorithms $SM^{i}$, with $i=0, 1, 2$. For comparison we plot the green staircase, $K_{\text{max}}$, and the analytic formula $R(N)$ (dashed green line). 
This figure shows the improvements that result from the (considerable) extra computational cost of the latter two algorithms.  Both the blue points of   $SM^{1}$ and the orange points of $SM^{2}$ reflect the staircase of $K_{\text{max}}$. Even their error bars reflect the rapid increase of the number of the larger maximum cliques after each jump in the staircase.  The red points of $SM^{0}$, although significantly greater than $\log_2N$, do not show any staircase pattern.   Each red point is the average over $2000$ random E-R graphs, each blue point the average over $500$ random E-R graphs, and each coral point is the average over $100$ random E-R graphs. We have used a uniform random number generator with extremely long period (WELL1024) \cite{panneton2006improved}.  The $SM^{2}$ results track the staircase closely up to $N = 4000$, the largest size seen in the DIMACS study, while the $SM^{1}$ results fall about $1$ site below the staircase at the end of this range.

\begin{figure}[t]
\centering
\subfloat[]{\label{fig:mdleft}{\includegraphics[width=1\columnwidth, keepaspectratio=true, angle=0]{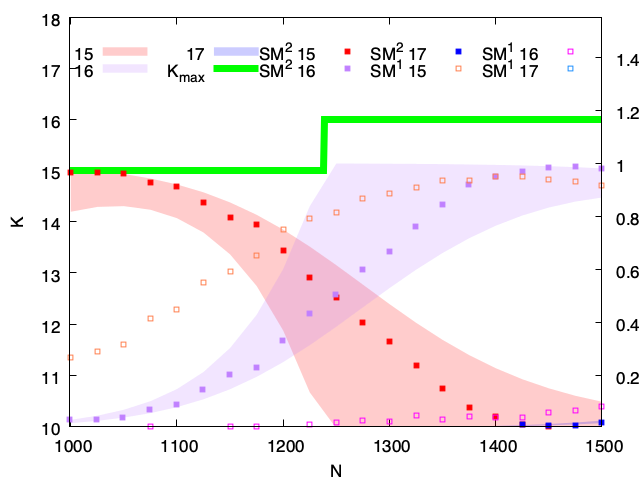}}}\hfill
\subfloat[]{\label{fig:mdright}{\includegraphics[width=1\columnwidth, keepaspectratio=true, angle=0]{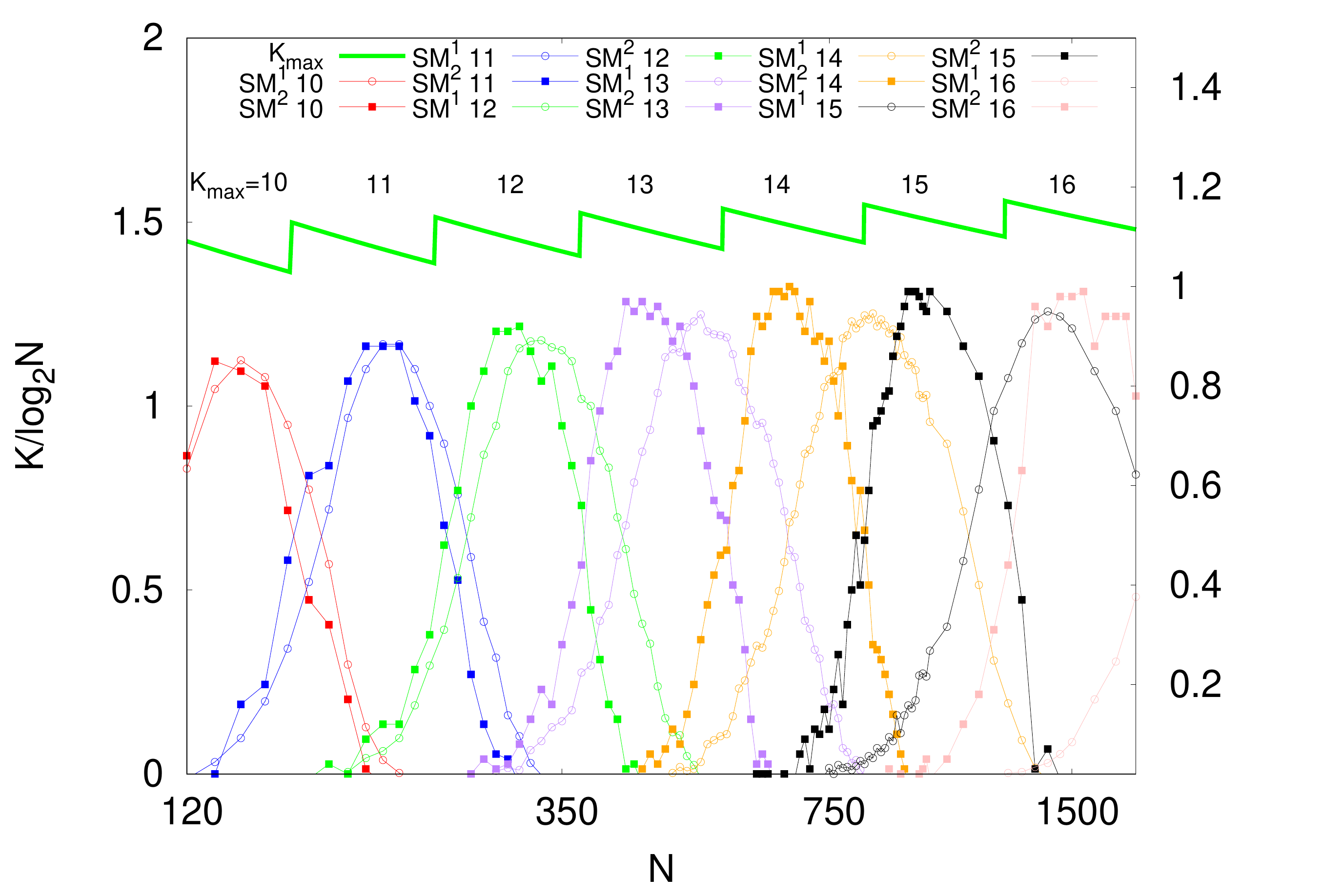}}}
\caption{(\textbf{a}) Across a single step in $K_{\text{max}}$, this figure compares the fraction of graphs predicted to have each value of $K_{\text{max}}$ from equation (\ref{eq::2}) with the experimental probability obtained using $SM^1$ and $SM^2$. The left $y$-axis shows the clique number, i.e. the largest clique size $K_{\text{max}}$ as a function of $N$. The right $y$-axis indicates the fraction of $K_{\text{max}}-1$, $K_{\text{max}}$, $K_{\text{max}}+1$, obtained from the experiments with $SM^1$ and $SM^2$. Each fraction has been computed on a population of $500$ random E-R graphs $G(N,p=0.5)$. The regions coloured in red, purple and blue identify the theoretical probability obtained by equation (\ref{eq::2}). The filled square points describe the fraction of $K_{\text{max}}-1$, $K_{\text{max}}$, $K_{\text{max}}+1$ (red, purple and blue respectively) obtained from experiments with $SM^2$, the empty squares the same for $SM^1$. (\textbf{b}) Here  we follow the fraction of graphs found by $SM^1$ (open-circles) and $SM^2$ (solid-squares) from $K_{\text{max}} = 10$ to $K_{\text{max}} = 16$.  The fraction seen for each value of $K_{\text{max}}$ receives a different color.}
\label{figSM1SM2}
\end{figure}

Figs. \ref{figSM1SM2}(\textbf{a}) and \ref{figSM1SM2}(\textbf{b}) give a more detailed comparison of the two algorithms.  Fig. \ref{figSM1SM2}(\textbf{a}) compares the predicted fraction of random E-R graphs having a maximal clique size with the experimental results obtained with the two algorithms around the step from $K_{\text{max}} = 15$ to 16. This corresponds to the region most often explored in the DIMACS studies.
The $SM^{2}$ algorithm remains within the bounds described by Matula.  The red, purple, and blue filled square points in the three predicted probability regions (red, purple, and blue, respectively) find acceptable fractions of $15$, $16$ and even $17$ sites cliques as $N$ is increased.  The $SM^{1}$ algorithm, shown by  orange, pink, and blue empty squares, falls short in all three probability regions, finding too many $15$'s, too few $16$'s and no $17$'s. 

If we expand the scale, covering steps from $K_{\text{max}} = 10$ to $K_{\text{max}} = 16$, we can see how the $SM^1$ results, which track closely with the $SM^2$ for $K_{\text{max}} = 10$ and $11$, fall behind as $N$ increases to give $K_{\text{max}}$ values of $12-14$.  For $N>1000$, $SM^1$ no longer reaches cliques with the true value of $K_{\text{max}}$.  $SM^2$, however, not only produces cliques with the step value of $K_{\text{max}}$, it correctly predicts a $50\%$ admixture of the two values of $K_{\text{max}}$ on which the problem is concentrated at each step where $K_{\text{max}}$ changes.

These two algorithms are very expensive. We could only analyze rather small random graphs. Next we consider less costly algorithms, which allow us to explore much larger graphs. These give results lying between $SM^{0}$ and $SM^{2}$and still reflect the staircase character of the underlying problem.

We reverse the order of operations made by the class of algorithm $SM^{i}$, with $i=1, 2, ..$. Instead of running $SM^{0}$ for each pair, or triangle, or tetrahedron (etc.) in the original graph, we run $SM^{i}$, with $i=1, 2, ..$ fixed, but only on the sites found within one solution given by $SM^{0}$.  $SM^0$ will return a clique $C$ of size $|C|$. On this solution we run $SM^i$, i.e. we select all the possible ${|C|\choose i}$ complete subgraphs  in the clique $C$, and, on each of them, denoted $C^{'}$, we run a restricted $SM^0$. In other words, we run  $SM^0$ on the graph $G^{''}$ induced by all sites $v \in C^{'} \cup N(C^{'})$. This simple algorithm, that we call $SM^0 \to SM^i$, will run in a time bounded by $\mathcal O(N^2 \ln N)$.
 

	
\begin{figure}
\includegraphics[width=1\columnwidth, keepaspectratio=true, angle=0]{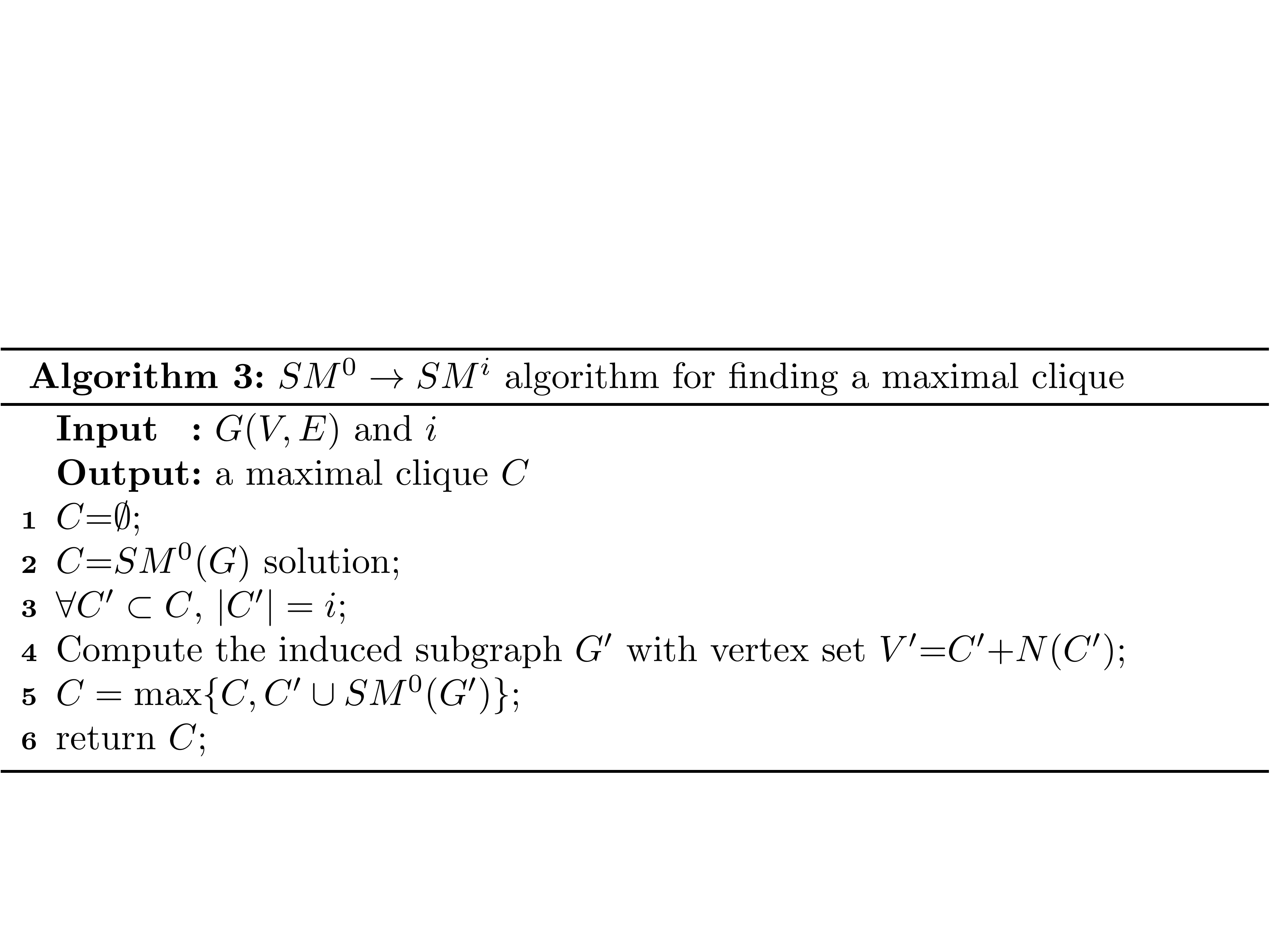}
\label{algo3}
\end{figure}

\begin{figure}
\centering
\includegraphics[width=1\columnwidth, keepaspectratio=true, angle=0]{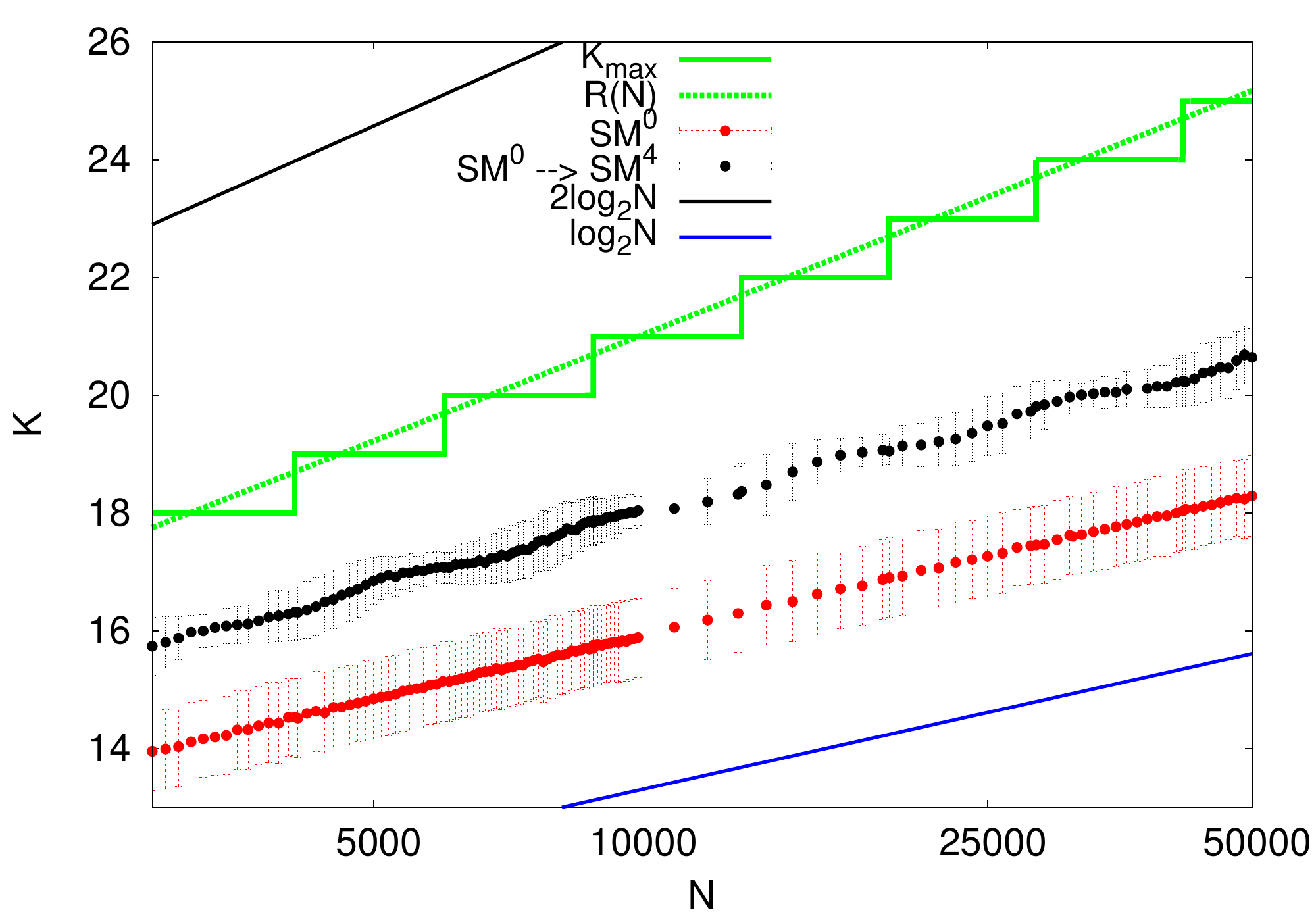}
\caption{The size of the maximal cliques found on E-R graphs $G(N, p=0.5)$, using the algorithm $SM^{0}\to SM^{4} $.  The green staircase indicates $K_{\text{max}}$, while the dashed green straight line represents  $R(N)$. The black and the blue straight lines describe the values of $2\log_2 N$ and $\log_2 N$, respectively. Red circles describe the maximal clique size mean obtained by $SM^{0}$, averaging over 2000 random E-R graphs, as a function of the order of the graph. Black circles are the mean maximal clique size obtained by $SM^{0}\to SM^{4}$, averaging over 500 random E-R graphs. }
\label{figSM0toSM4}
\end{figure}

As an example, we show in Fig. \ref{figSM0toSM4} the results of the algorithm $SM^{0}\to SM^{i}$, with $i$ fixed to $4$,  compared to  $SM^0$ in the range of $N$ $[2800:50000]$. We analyze ${|C|\choose 4}$ graphs of order approximatively $N/16$. The combined algorithm $SM^{0}\to SM^{4}$ always finds a maximal clique bigger than those given by $SM^0$ alone. Moreover the combined algorithm reproduces the wiggling behaviour due to the discrete steps in $K_{\text{max}}$ in a time bounded by  $\mathcal{O}(N^2 \ln N)$, while $SM^{0}$, used alone, does not.

The improved results of the combined algorithm $SM^{0} \to SM^{i}$, with fixed ${i}$, suggests to iterate the procedure. First we run $SM^{i}$, with fixed $i$, on the clique returned by $SM^{0}$. If the clique returned by the algorithm is bigger than the one that is used for running $SM^{i}$, then we use the new clique as a starting point where $SM^{i}$ will be run again. The algorithm stops when the size of the clique no longer increases. The complexity of the algorithm therefore is  $\mathcal{O}(tN^2 \ln N)$, where $t$ is the number of times we find a clique which is bigger than the previous one. We call, thus, this new algorithm $SM^{0}\to \text{iter}[SM^{i}]$.

\begin{figure}
\includegraphics[width=1\columnwidth, keepaspectratio=true, angle=0]{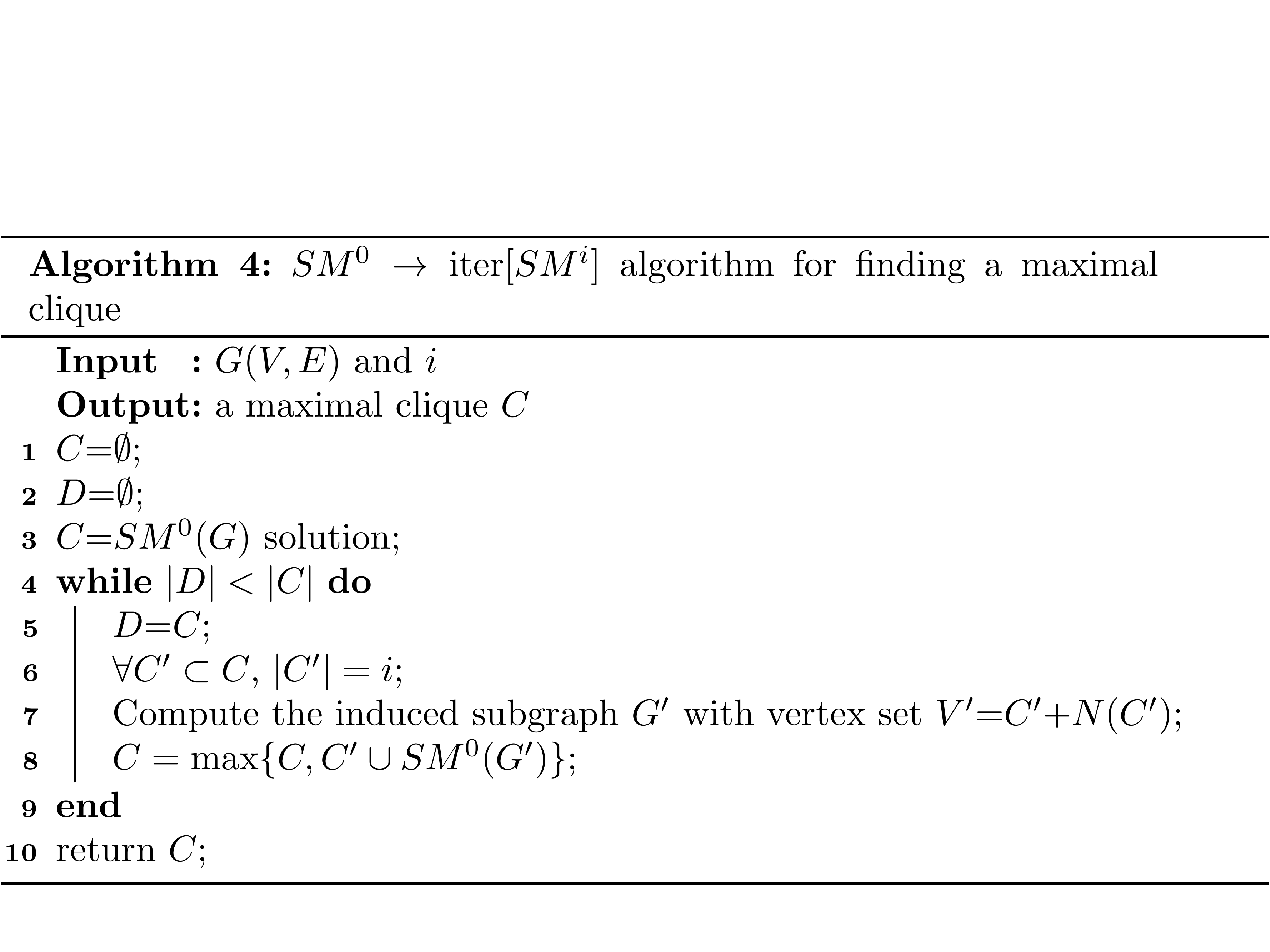}
\label{algo4}
\end{figure}

\begin{figure}
\centering
\includegraphics[width=1\columnwidth, keepaspectratio=true, angle=0]{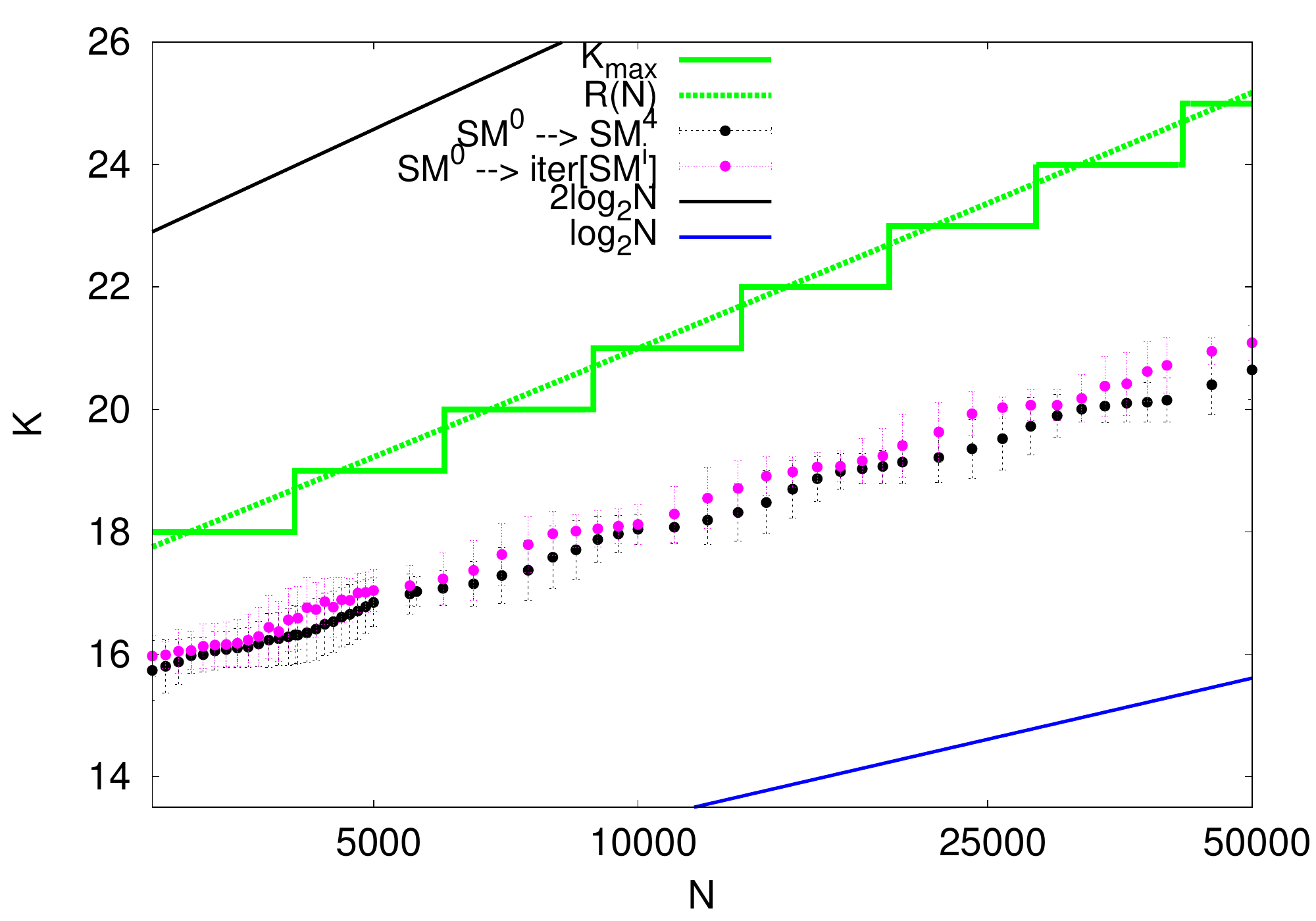}
\caption{Maximal clique on E-R graphs $G(N, p=0.5)$, using the algorithm $SM^{0}\to \text{iter}[SM^{i}] $.  $K_{\text{max}}$ is represented by the green staircase. The dashed green straight line is $R(N)$. The black and the blue straight lines describe the values of $2\log_2 N$ and $\log_2 N$, respectively. Black circles are the mean maximal clique size  obtained by $SM^{0}\to SM^{4}$, averaging over 500 random E-R graphs, while magenta circles are the mean maximal clique size obtained by $SM^{0}\to \text{iter}[SM^{i}] $, using the values of $i$ in Table (\ref{Tabi}).}
\label{figSMiterwhichi}
\end{figure}

We present in Fig. \ref{figSMiterwhichi} the results of $SM^{0}\to \text{iter}[SM^{i}]$, over the range of $N$ from $2000$ to $50000$, comparing them with the results of $SM^{0}\to SM^{4}$.  We use different $i$ in different ranges of $N$, determining their values by experiment. As $N$ increases we have to increase the number of sites kept for the iteration in order to get a bigger complete subgraph at the end of the process.  

\begin{figure}
\centering
\includegraphics[width=1\columnwidth, keepaspectratio=true, angle=0]{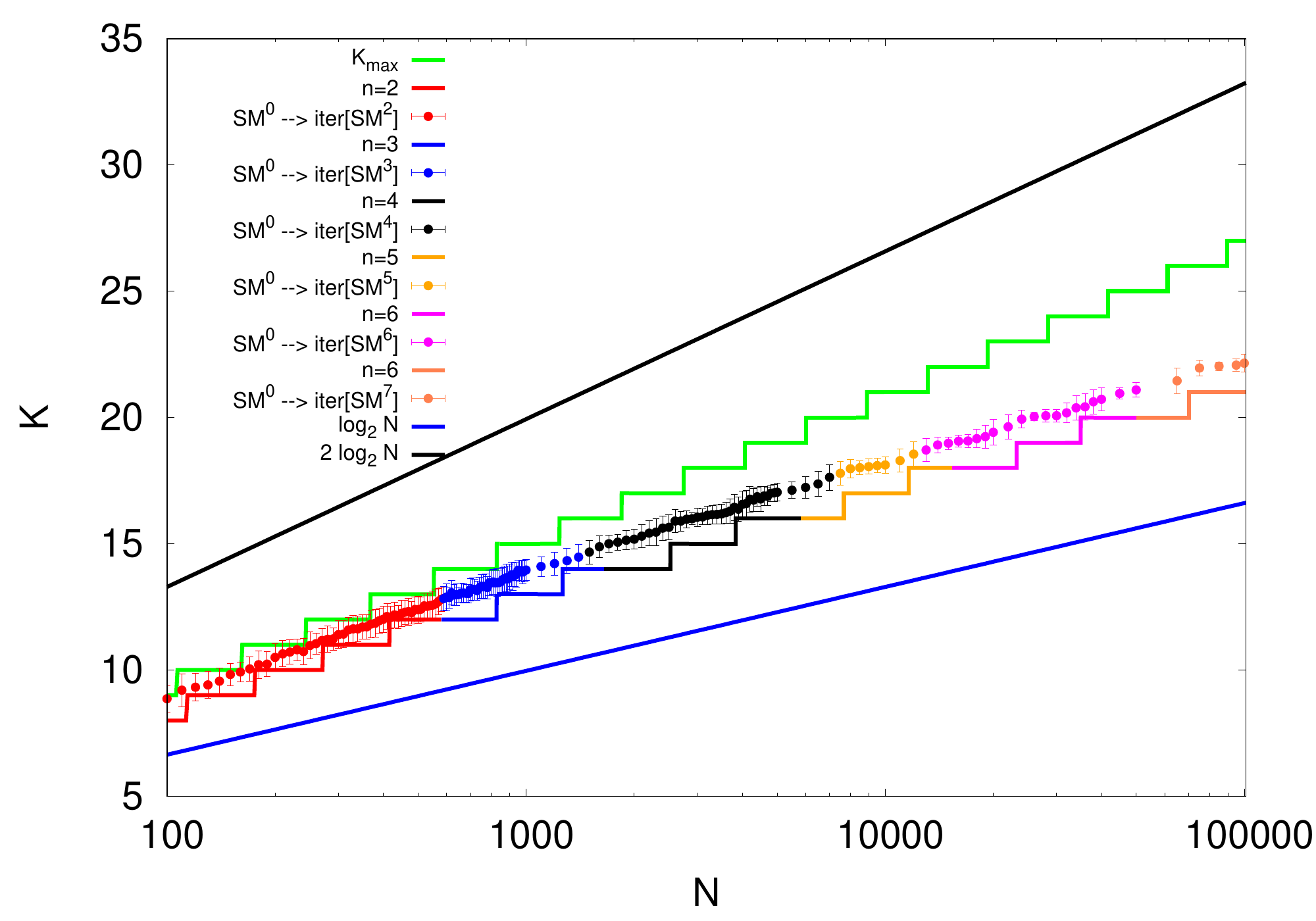}
\caption{Maximal clique sizes found on E-R graphs $G(N, p=0.5)$.  The green staircase is $K_{\text{max}}$, as in previous figures. Data points are the                 mean maximal clique size obtained by $SM^{0}\to \text{iter}[SM^{i}] $, using values of $i$ given by Table \ref{Tabi}. The multi-coloured staircase represents the expected maximum values of completed subgraphs conditioned by the fact that we are starting with an arbitrary  complete subgraph of size $i$.}
\label{figSMiterwhichi1}
\end{figure}

The values of $i$ selected are given in the following table: 
\begin{equation}
\label{Tabi} 
 \begin{tabular}{||c | c ||} 
 \hline
$N$ & $i$ \\ [0.5ex] 
 \hline
 100-589 & 2 \\ 
 \hline
 590-1499 & 3\\
 \hline
1500-7499 & 4\\
 \hline
7500-12999 & 5\\
 \hline
13000-64999 & 6\\
 \hline
 65000-100000 & 7\\ [1ex] 
 \hline
\end{tabular}
\end{equation}
 
Fig. \ref{figSMiterwhichi1} shows the results of experiments with $SM^{0}\to \text{iter}[SM^{i}] $, with $i$ fixed in the range given by Tab. \ref{Tabi}. They fall between two \textit{staircases}.  The upper one is $K_{\text{max}}$, as before, and the lower one is the max clique size predicted by the first moment bound if we begin with a randomly selected clique of size $i$.  The coloured staircase curve is given by \cite{feller1} the smallest value of $K$ for which : 
\begin{equation}
\label{eqcsc}
{N - i \choose K -i }p^{{K\choose 2}-{i\choose 2}} \geq 1.
\end{equation}
This implies that the subgraphs we have selected as a basis for our iteration are much better than average, compared with the very large number of starting subgraphs that a full $SM^i$ would have required.

\subsection{Other concentrations of links}

The same estimates of clique sizes and techniques for finding them will of course be extended to work at values of $p$ other than $0.5$, but finite size effects must be carefully considered. For general values of $p$, the extrapolation based on Stirling's approximation extends to \cite{bollobas1976cliques,matula1970complete,matula1976largest,frieze1990independence}:
 \begin{equation}
 R(N,p)= 2\log_{1/p} N-2\log_{1/p} \log_{1/p} N+2 \log_{1/p} e/2 +1
 \end{equation}
 
\begin{figure}[t]
\centering
\subfloat[]{\label{fig:mdleft}{\includegraphics[width=1\columnwidth, keepaspectratio=true, angle=0]{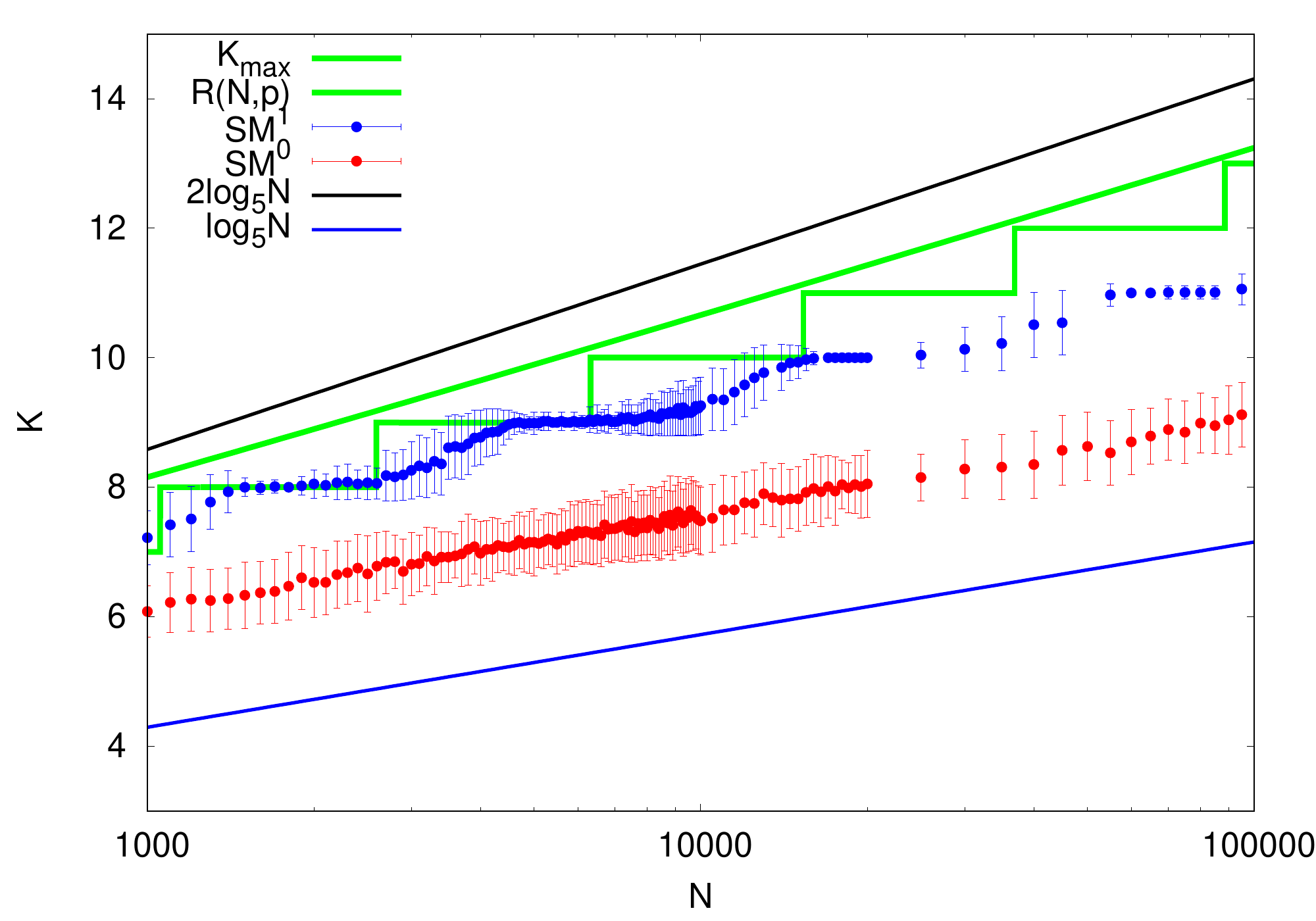}}}\hfill
\subfloat[]{\label{fig:mdright}{\includegraphics[width=1\columnwidth, keepaspectratio=true, angle=0]{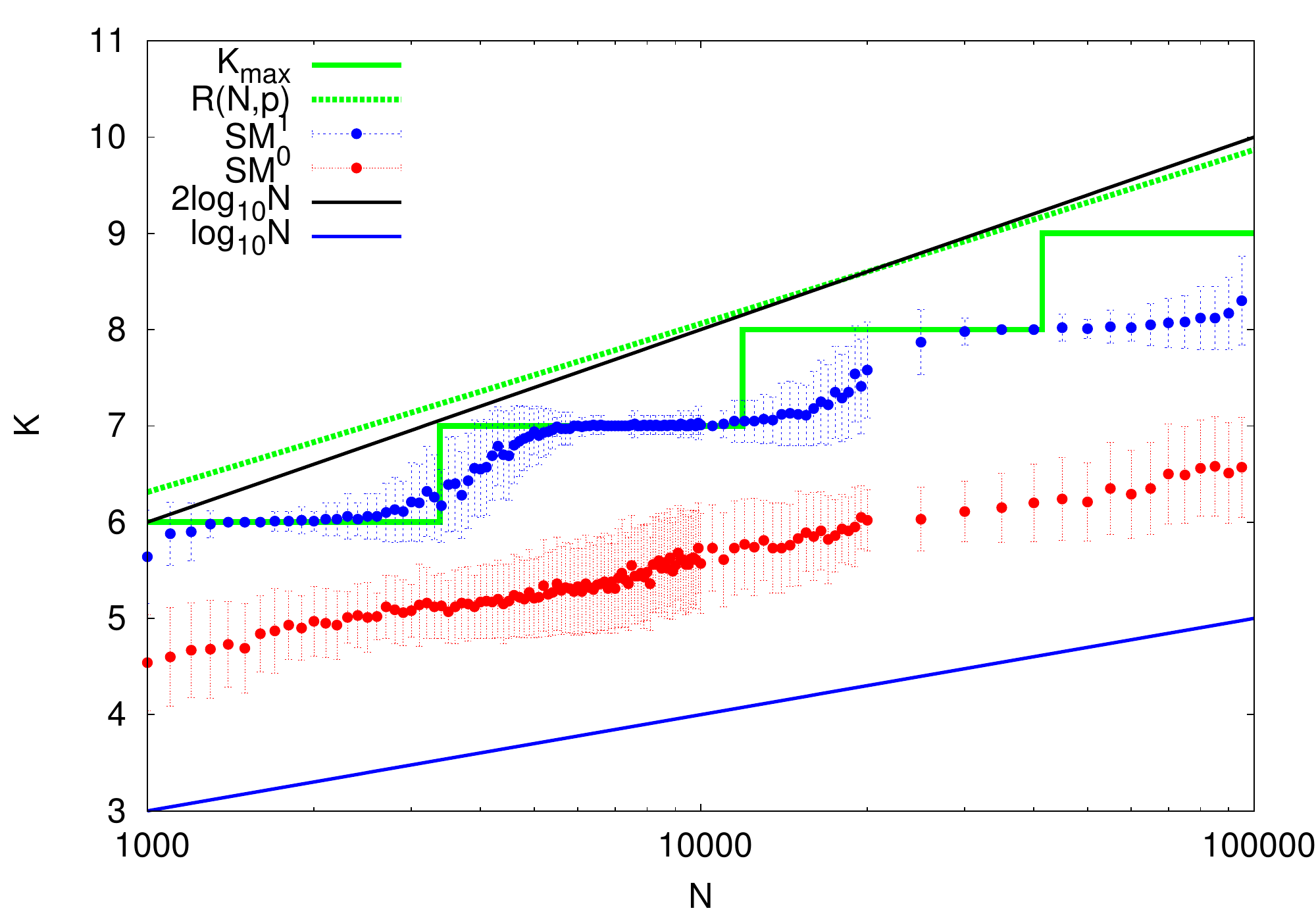}}}
\caption{Algorithms $SM^{0}$ and $SM^1$ applied to $G(N,p)$ graphs with $p = 0.2$ (\textbf{a}) and $0.1$ (\textbf{b}). Results for $SM^0$ and $SM^1$  were obtained for values of $N$ up to $95000$ in both cases.  The error bars indicate standard deviations over samples of $100$ graphs at each data point.}
\label{figsmallp}
\end{figure}

Although when $p = 0.5$, $R(N,p)$ can be used for all values of N that we have considered, much larger or smaller values of $p$ require larger values of $N$ for its value to remain between the two values of $K$ upon which the maximum clique sizes are concentrated.  When  $p$ is much larger than $0.5$, $R(N,p)$ may fail to cross through the rising portion of each $K_{\text{max}}$ step.  For $N = 10^3$, $R(N,p)$ falls below the steps when $p>0.65$.  For $N = 10^6$, $R(N,p)$ remains a useful guide only up to $p \sim 0.7$.  The same problem occurs at small values of $p$, with the value of $R(N,p)$ exceeding the actual expected values of $K_{\text{max}}$ at sufficiently low $p$ and insufficiently large $N$.  Thus $R(N,p)$ rises above the steps when $N > 10^3$ and $p < 0.35$, while for $N = 10^6$, $R(N,p)$ continues to cross through the rise between steps down to about $p \sim 0.15$.  Nonetheless, our algorithms $SM^i$ still work at smaller $p$, as the next two figures show.  In Fig \ref{figsmallp} (\textbf{a}) we show the results of running $SM^0$ and $SM^1$ at $p = 0.2$ and in Fig \ref{figsmallp}(\textbf{b}) we show the same algorithms applied to $p = 0.1$.  The solid lines in this figure indicate the simple estimates $\log_{1/p} N$ and $2 \log_{1/p} N$, respectively. In both cases, $SM^0$ provides cliques at about $1.5\, \log_{1/p} N$, or $1.5 \times$ the naive greedy result, while $SM^1$ captures the oscillation of the steps up to well above $N = 10^4$ and runs fast enough to have given us data averaged over $100$ graphs at sizes up to $N = 95000$ at $p=0.2$ and $p = 0.1$.  

We performed the same tests to see how well $SM^1$ reproduced the distribution of graphs with values of $K_{\text{max}}$ from just below and just above a step that were presented in Fig \ref{figSM1SM2} (\textbf{a}) and \ref{figSM1SM2} (\textbf{b})(for $p = 0.5$) on our two cases at $p = 0.2$ and $0.1$, using Matula's first and weighted second moment calculations to provide upper and lower bounds on the fraction of graphs with each value of $K_{\text{max}}$.  As was the case at $p = 0.5$, at each step, the fraction of graphs having at least one clique with the new, higher value of $K_{\text{max}}$ is asymptotically half.  

At $p = 0.2$, we tested the steps from $K_{\text{max}}=8$ to $9$ and from $9$ to $10$.  Although at these steps, and the following one, the average value of $K_{\text{max}}$ reported by $SM^{1}$ follows the lower part of each step, at the step from $8$ to $9$, $SM^1$ found $K_{\text{max}} = 9$ in only $20\%$ of the graphs considered, reaching $80\%$ in the middle of the step above.  At the step from $9$ to $10$, $SM^1$ found $K_{\text{max}} = 10$ in only $5\%$ of the graphs, reaching only $10\%$ at the middle of the step.  The case $p = 0.1$ was easier for these values of $N$ and $SM^1$.  As is apparent in Fig. \ref{figsmallp} (\textbf{b}), where the variation in the results spans the step height from $K_{\text{max}} = 6$ to $7$, we found the larger value of $K_{\text{max}}$ in roughly $40\%$ of the graphs, reaching $80\%$ at the middle of the following step.  But at the step from $K_{\text{max}} = 7$ to $8$, occurring at roughly $N = 12000$, $SM^1$ observed the larger value of $K_{\text{max}}$ in only about $6\%$ of the graphs, increasing to less than $20\%$ at the middle of the following step.  As $N$ increased further, $SM^1$ reached cliques of size $8$, but increased only slowly thereafter.

\begin{figure}
\centering
\includegraphics[width=1\columnwidth, keepaspectratio=true, angle=0]{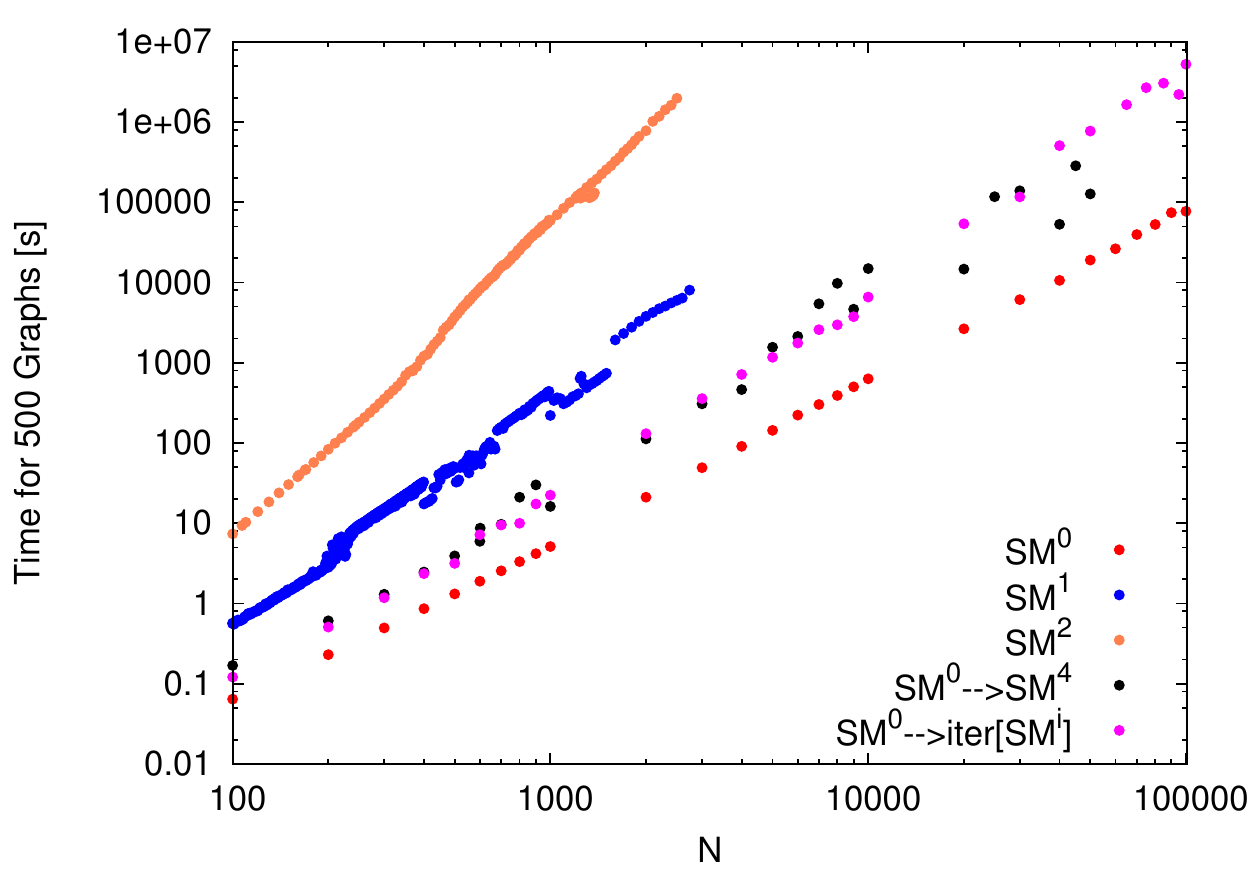}
\caption{The picture shows the cost of the algorithms in seconds $[s]$ for solving $500$ random E-R graphs of order $N$ and with edge density equal to $p=0.5$. }
\label{costofalgo}
\end{figure}

Fig. \ref{costofalgo} shows, finally, the cost of algorithms for $500$ random E-R graphs.  From the slopes we see  that red points, i.e. $SM^0$, are proportional to $\mathcal{O}(N^2)$, blue points, i.e. $SM^1$, to $\mathcal{O}(N^3)$ and the coral points to $\mathcal{O}(N^4)$, as expected. Black and magenta points, which identify the cost of $SM^{0}\to SM^{4}$ and $SM^{0}\to \text{iter}[SM^{i}] $ are bounded by polynomial functions of order $\mathcal{O}(N^2 \ln N)$  and $\mathcal{O}(t N^2 \ln N)$, respectively.

\begin{figure}
\centering
\includegraphics[width=1\columnwidth, keepaspectratio=true, angle=0]{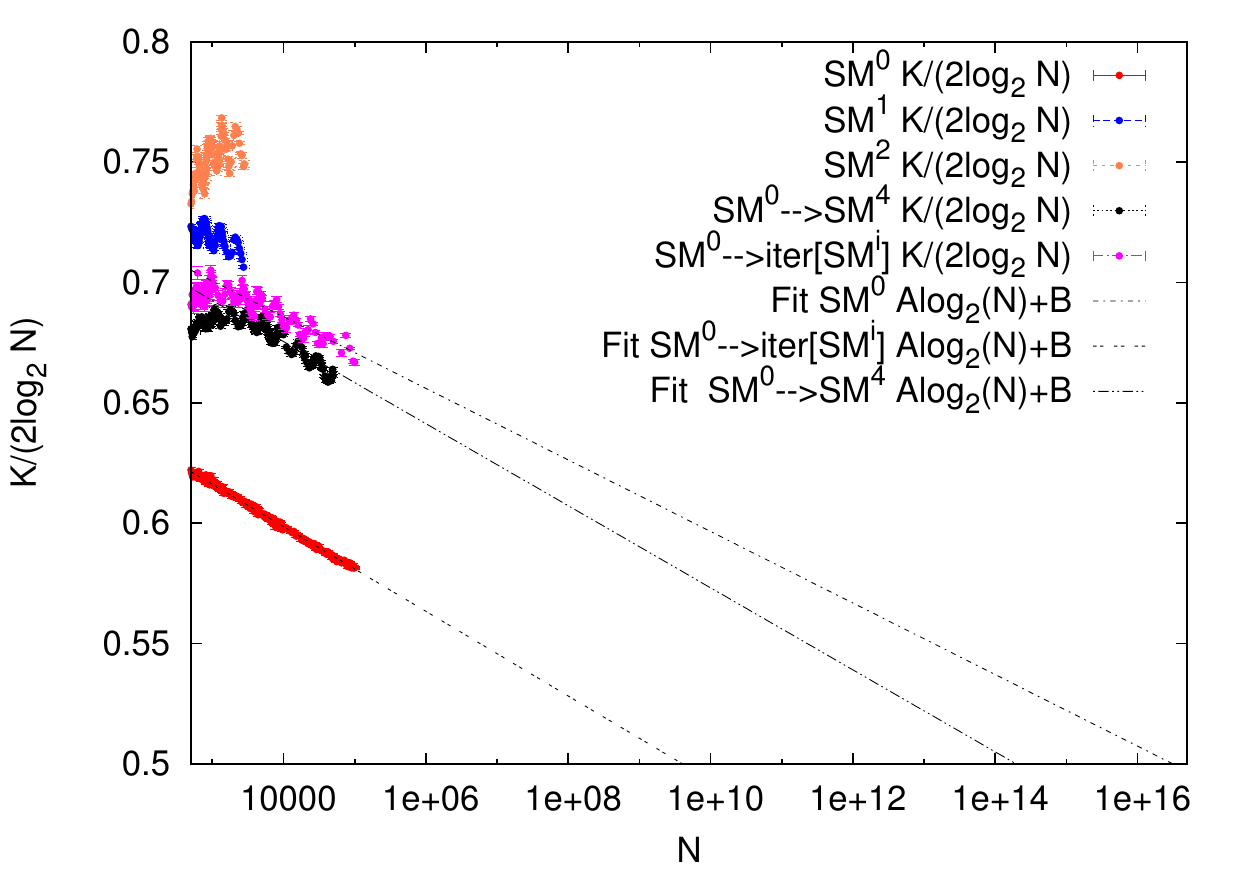}
\caption{The picture shows the inferred limit of the algorithms. Red points, i.e. $SM^0$, extrapolates at $5\times10^9$, with values of $A=-0.005$ and $B=0.668$, black and magenta points, which identify $SM^{0}\to SM^{4}$ and $SM^{0}\to \text{iter}[SM^{i}] $, extrapolate at $10^{14}$, with values of $A=-0.005$ and $B=0.743$, and $2\times10^{16}$,with values of $A=-0.004$ and $B=0.745$, respectively. Coral and blue points, $SM^2$ and $SM^1$ respectively, are not extrapolated.}
\label{Jerummconf}
\end{figure}

The last picture that we present here, i.e. Fig. \ref{Jerummconf}, displays the inferred limit of the algorithms. In this picture we plot the normalised maximal clique, i.e. $K/2\,\log_2 N$, as function of $N$, with the aim of determining the range of values of $N$ where the results of our polynomial cost algorithms exceed $\log_2 N$, thus meeting the \textit{challenges}.

We observe that all the linear and quasi-linear algorithms presented in this work succeed in finding a maximal clique of size at least $\log_2(N)$  until $N$, the order of the graph becomes bigger than or equal to  $5\times10^9$ if we use $SM^0$, $N \approx 10^{14}$ if we use $SM^0 \to SM^4$ and $N \approx 2\times10^{16}$ if we use $SM^{0}\to \text{iter}[SM^{i}] $.  The more elaborate cubic and quartic algorithms are not extrapolated, as we could not carry them out on large enough samples. 

The feature that makes this problem so hard is that at each step on our staircase, there are multiple cliques of size $K_{\text{max}}$, their number increasing from a few near the step edge, to a fairly large number as we approach the next step. There are still larger numbers of size less than $K_{\text{max}}$.  We can make a rough estimate of the number of cliques of size $K_{\text{max}}$ that a randomly selected site lies in by multiplying the expected number of such cliques by their size, and dividing by $N$. This is justified since calculating the expected number of pairs of such large cliques overlapping by a fixed number of sites, we find that these are dominated by overlaps of zero or one site.  This estimate, an upper bound on the probability that a site is part of a maximum clique is zero at the start of each step (when such cliques are rare) and increases to between $5$ or $6$ by the end of the step, and is greater than $1$ for only about the last $20\%$ of each step.  This is why $SM^1$ or $SM^2$, by selecting the best starting points, can give results better than the naive random start of $SM^0$.  

Cliques of size $\log_2 N$ exist in large numbers and overlap strongly. The fraction of them that will be entirely contained in a single clique of the largest size is vanishingly small. Thus most cliques that we find by a simple greedy construction consist of members or seeds of several to many of the larger cliques.  
Greedy search to grow a small clique which contains the seeds of several maximal cliques will eliminate from the frontier half of the remaining parts of each of its targets, and find it impossible to reach more than a fraction of $K_{\text{max}}$.  Our iterative procedure to search for the best small subsets from the largest cliques reached in a single greedy $SM^0$ search is an attempt to overcome this.  But since the cliques that can be formed before the dynamical threshold are strongly overlapping at all degrees, mistakes will still be made.  We see this in Fig \ref{figSMiterwhichi1}, where the lower set of steps show how much smaller are the cliques that one can expect to construct if they are forced to start with a random completely connected subgraph.  Although we choose the best such subset that we can find within a restricted space, and obtain a significant improvement, we still fall well short of the true limit $K_{\text{max}}$. 

Much discussion of the difficulty of searching large complex systems with many parameters over which to optimize uses language such as "gradient descent in a random potential", and "avoiding local minima at higher energies" than the goal.  The search for a maximal clique is perhaps even more difficult, but the difficulty seems to be a total lack of information with which to choose between the possible paths to different solutions. Early incorporation of the seeds of multiple, incompatible solutions  produces interference as the seeds of each partial solution rule out the remainder of the others. 

This has caused several workers to shift attention to a problem in which there is only one "planted" clique to be sought, which can be identified by having a size larger than is expected to result from random graph processes. The hope is that having a single target will make the the search to recover this objective more effective.  Many authors  \cite{jerrum1992large,alon1998finding,dekel2014finding,deshpande2015finding}  have introduced such "planted" or hidden solutions in order to test more powerful methods of discovering them.  In the next section, we consider some of these methods, present some novel extensions, and compare them to our greedy search techniques.

\section{Hidden Clique}

To perform computer experiments, it is conventional to use the first $K_{HC}$ sites as the hidden subset, which makes it easy to observe the success or failure of oblivious algorithms as well as those to which we will give \textit{hints}.  Hints are a quite reasonable part of the hidden clique problem, as many practical problems in information retrieval take the form \textit{"find a community that closely resembles or is linked strongly to one or more exemplars"}. Having a labelled hidden clique permits experiments in which we can easily see how many of the hidden clique's sites would have been discovered by a particular search strategy.

We construct the hidden clique in one of two ways.  The first is simply to restore all the missing links among the first $K_{HC}$ sites.  This has the drawback that those sites will have more neighbors than average, and might be discovered by exploiting this fact. In fact, the upper limit to interesting hidden clique sizes was pointed out by Ku{\v{c}}era \cite{kuvcera1995expected}, who showed that a clique of size $C \sqrt{N \ln N}$ for a sufficiently large $C$ will consist of the sites with largest number of neighbors, and thus can be found by $SM^0$.

The second method is to move links around within the random graph in such a way that after the hidden clique is constructed, each site will have the same number of links that it had before.  To do this, before we add a link between sites $i$ and $j$ in the hidden clique, we select at random two sites, $k$ and $l$, which lie outside the clique. $ k$ must be a neighbor of $i$ and $l$ must be a neighbor of $j$.  If $k$ and $l$ are distinct and not neighbors, we create a new link between them, and remove the links between $i$ and $k$ and between $j$ and $l$.  If this fails we try the replacement again, still selecting sites $k$ and $l$ at random.  The result is a new graph with the same distribution of connectivities, as measured from the individual sites.  This sort of \textit{smoothing} of the planting of a hidden clique had been explored by  \cite{sanchis1994test}.  Several graphs prepared in this way are in the DIMACS portfolio, and appear to be more difficult to solve.

A stronger result, by Alon et al. \cite{alon1998finding} uses spectral methods, which we shall discuss in detail below, to show that a hidden clique, $C$, of cardinality  $|C|\ge 10 \sqrt{N}$ can be found with high probability, in polynomial time. Dekel et al. \cite{dekel2014finding} showed that with a linear ($\mathcal{O}( N^2)$, the number of links) algorithm the constant can be reduced to $1.261$. Finally, recent work of Deshpande and Montanari \cite{deshpande2015finding} has shown that Approximate Message Passing (\textbf{AMP}), a novel form of belief propagation, can also identify sites in the hidden clique.  This converges down to $\sqrt{N/e}$, where $e$ is Euler's constant.  No algorithm currently offers to find a clique of size less than $\sqrt{N/e}$  and bigger than $K_{\text{max}}$, in polynomial time, for arbitrary $N$. Each of these procedures identifies some, but perhaps not all of the planted clique sites, and requires some "\textit{cleanup}" steps to complete the identification of the whole clique. The cleanup procedures all require starting with either a subset of the hidden clique sites and finding sites elsewhere in the graph that link to all of them, or eliminating the sites in a possible mixed subset of valid and incorrect choices which do not extend as well, or doing both in some alternating process.  These can be proven to work if the starting point is nearly complete (hence Alon et al.'s $C = 10$ starting point).  We find experimentally, and discuss below, that a cleanup process can be effective given a much poorer starting point as well.

\subsection{Spectral Methods}

Alon et al.'s approach \cite{alon1998finding} requires the eigenvalues and eigenvectors of an $N\times N$ dimensional adjacency matrix for the graph.  While this is conventionally described as taking $\mathcal{O} (N^3)$ operations, modern linear algebra libraries, such as Armadillo \cite{sanderson2016armadillo}, based on LAPACK \cite{mlpack2018}, exploit the many cores available in a modern laptop, and achieve compute costs scaling as $\mathcal{O} (N^{2.5})$ over the range of $N$ we study experimentally.   This permits some interesting experiments, but first we need some definitions and derivations.

The adjacency matrix, $\mathbf{A}$, of our graph is actually a hybrid of two components, one for the random graph, and one for the hidden clique, each of which has known properties in isolation.

The adjacency matrix of a random graph is symmetric,  
\begin{equation*}
\begin{split}
&A_{ij}  =  \frac{1}{\sqrt{N}} { a_{ij}  },\\
& a_{ij} = a_{ji}  = 1 \text{ or } 0, \,\, a_{ii}=0,
\end{split}
\end{equation*}

\begin{figure}
\centering
\includegraphics[width=1\columnwidth, keepaspectratio=true, angle=0]{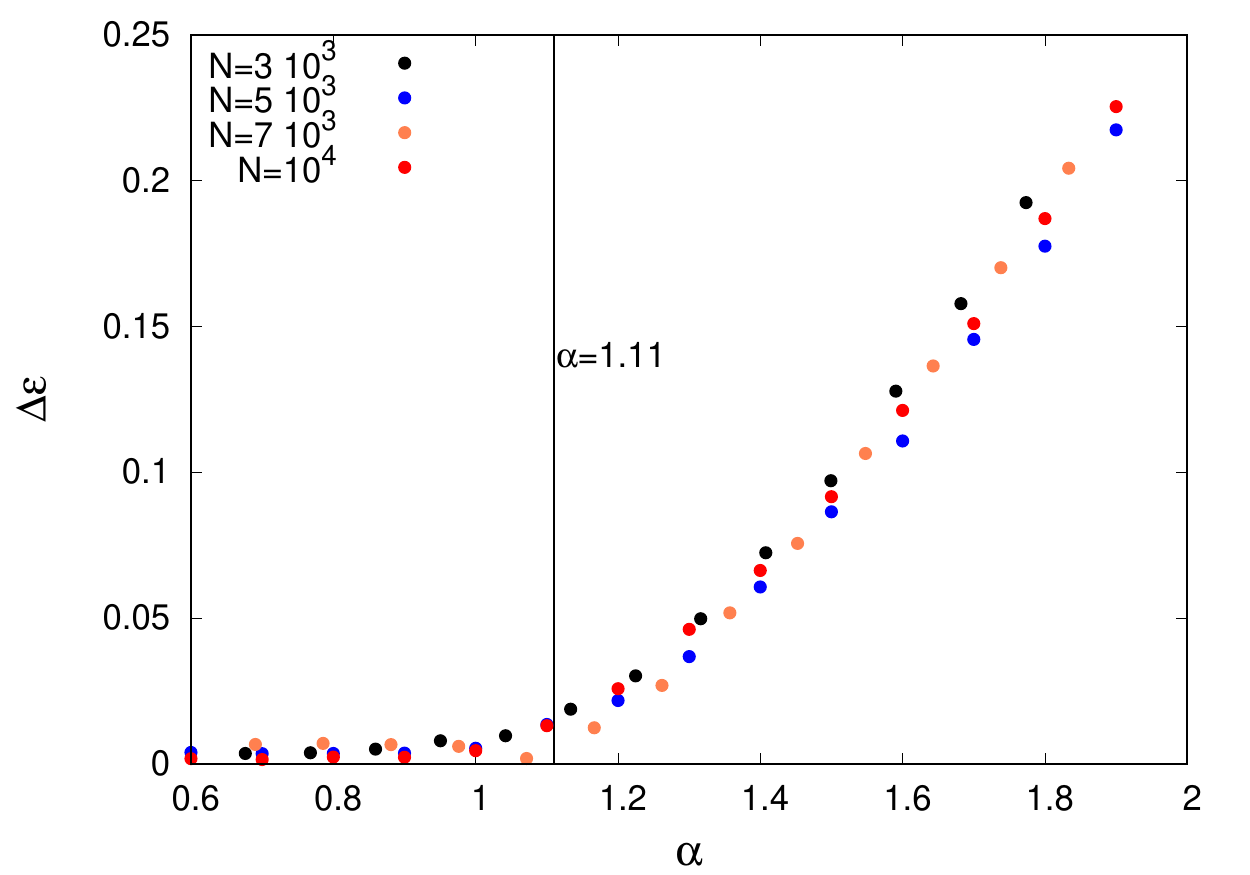}
\caption{The picture shows the energy gap value  as function of $\alpha=K_{HC}/\sqrt{N}$, for different values of $N$.  The gap appears  when $K_{HC}$ lies between $1.1$ and $1.2 \sqrt{N}$.}
\label{gap}
\end{figure}

has elements $a_{ij}$ which are $1$ on the links which are present in a given instance, and $0$ on the links which are absent, and $0$ on the diagonals.  One exceptional eigenstate of this matrix is the (nearly) uniform state, which for $p = 0.5$ has the eigenvalue $0.5 \sqrt{N}$.  The remaining eigenvalues, which are non-degenerate, fill a band from slightly less than $+1$ to slightly less than $-1$, and the spectral density in the limit of $N$ large is a semi-circle. The width of this band is set by the standard deviation of the off-diagonal elements of $\mathbf{A}$. Such matrices occur throughout physics. For example in the tight-binding model of electron motion in solids, the links are present between adjacent atoms in a solid, and represent the probability that an electron from one atom can hop to a similar orbital state on the next one.  In effect, $G(N,p)$ represents a sort of spherical (geometry-free) model of the energy band structure in a random system, with randomness coming from the missing links rather than from the random diagonal elements, or site energies, that could appear in a model of a 3D material.  We shall use this analogy later.

\begin{figure}
\centering
\includegraphics[width=1\columnwidth, keepaspectratio=true, angle=0]{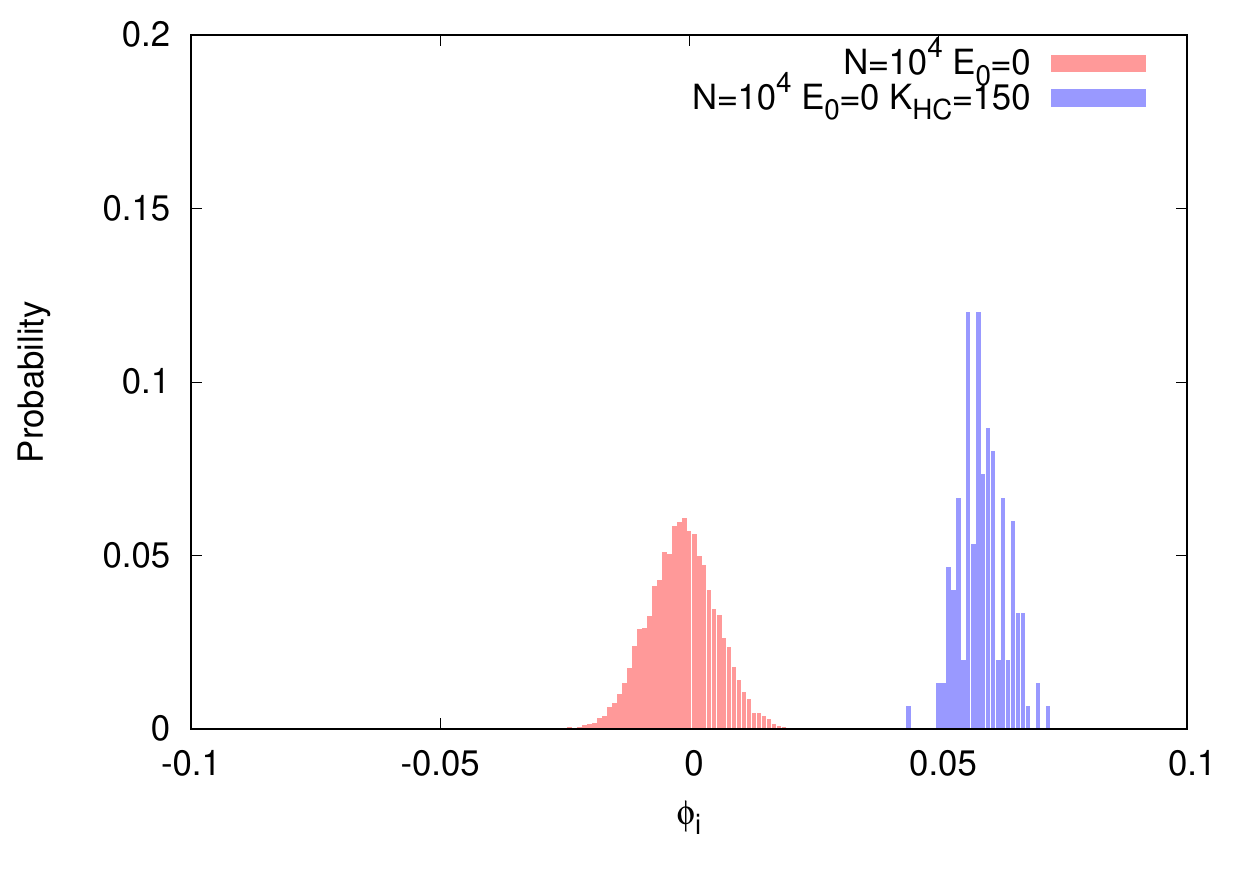}
\caption{The picture shows the distribution of the eigenvector components on the hidden clique sites for a case with $\alpha = 1.5$, blue histogram,  and the distribution of the components of that eigenvector on the other sites of the graph, red histogram.}
\label{distributioncomp}
\end{figure}


 \begin{figure}[t]
\centering
\subfloat[]{\label{fig:mdright}{\includegraphics[width=1\columnwidth, keepaspectratio=true, angle=0]{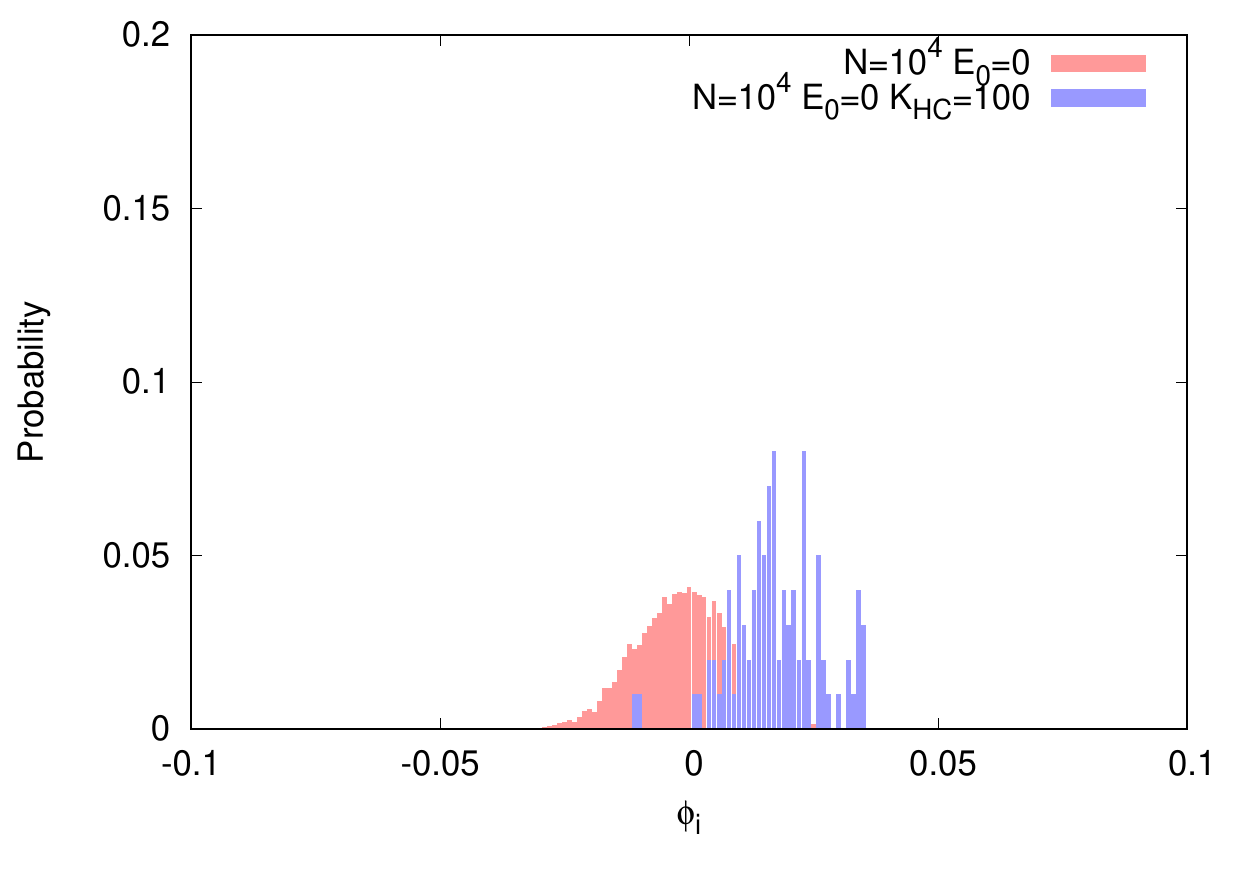}}}\hfill
\subfloat[]{\label{fig:mdright}{\includegraphics[width=1\columnwidth, keepaspectratio=true, angle=0]{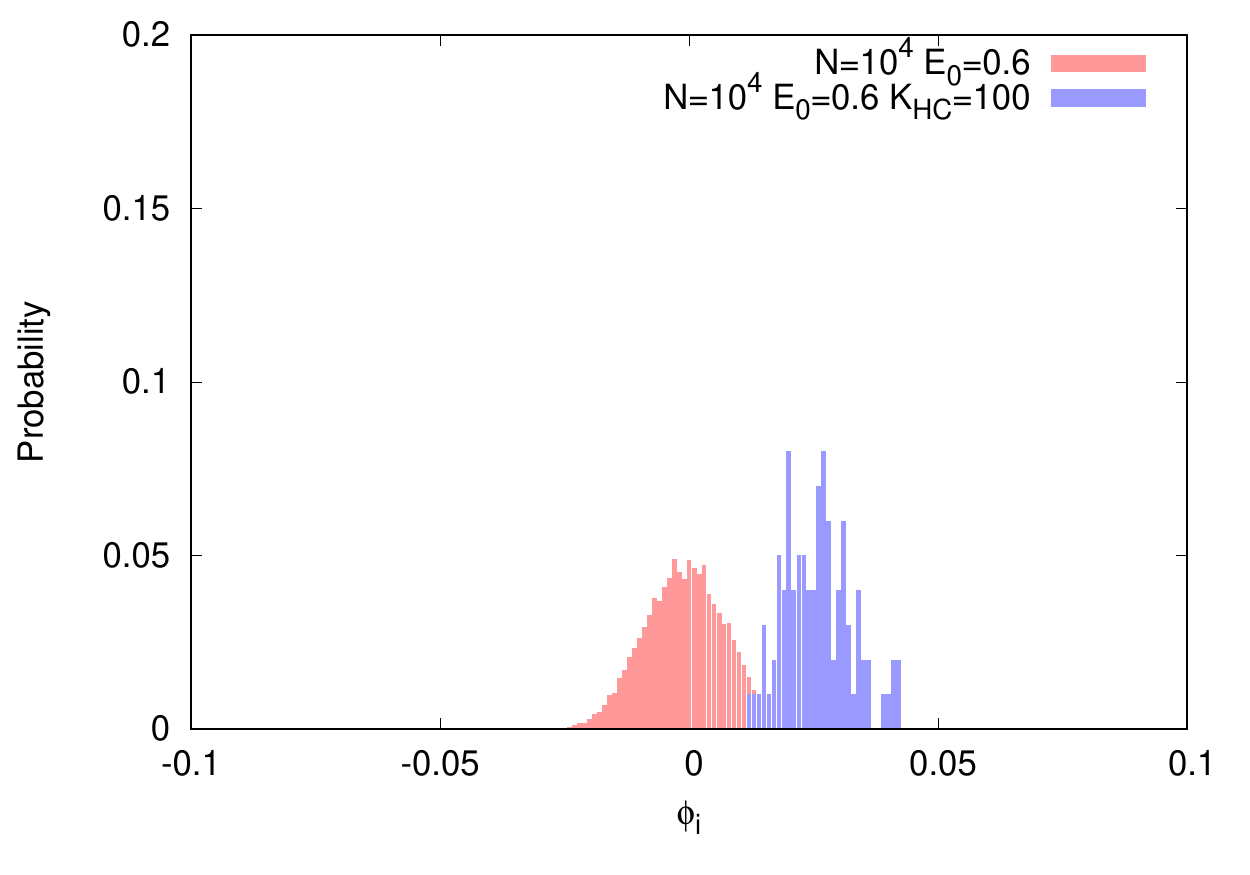}}}
\caption{The picture (\textbf{a}) shows the distribution of the eigenvector components on the hidden clique sites for a case with $\alpha = 1.0$, blue histogram,  and the distribution of the components of that eigenvector on the other sites of the graph, red histogram, when $E_0=0$, while the picture (\textbf{b}) shows the distribution of the eigenvector components on the hidden clique sites for a case with $\alpha = 1.0$, blue histogram,  and the distribution of the components of that eigenvector on the other sites of the graph, red histogram, when $E_0=0.6$.}
\label{distributioncomp-100}
\end{figure}

\begin{figure}
\centering
\includegraphics[width=1\columnwidth, keepaspectratio=true, angle=0]{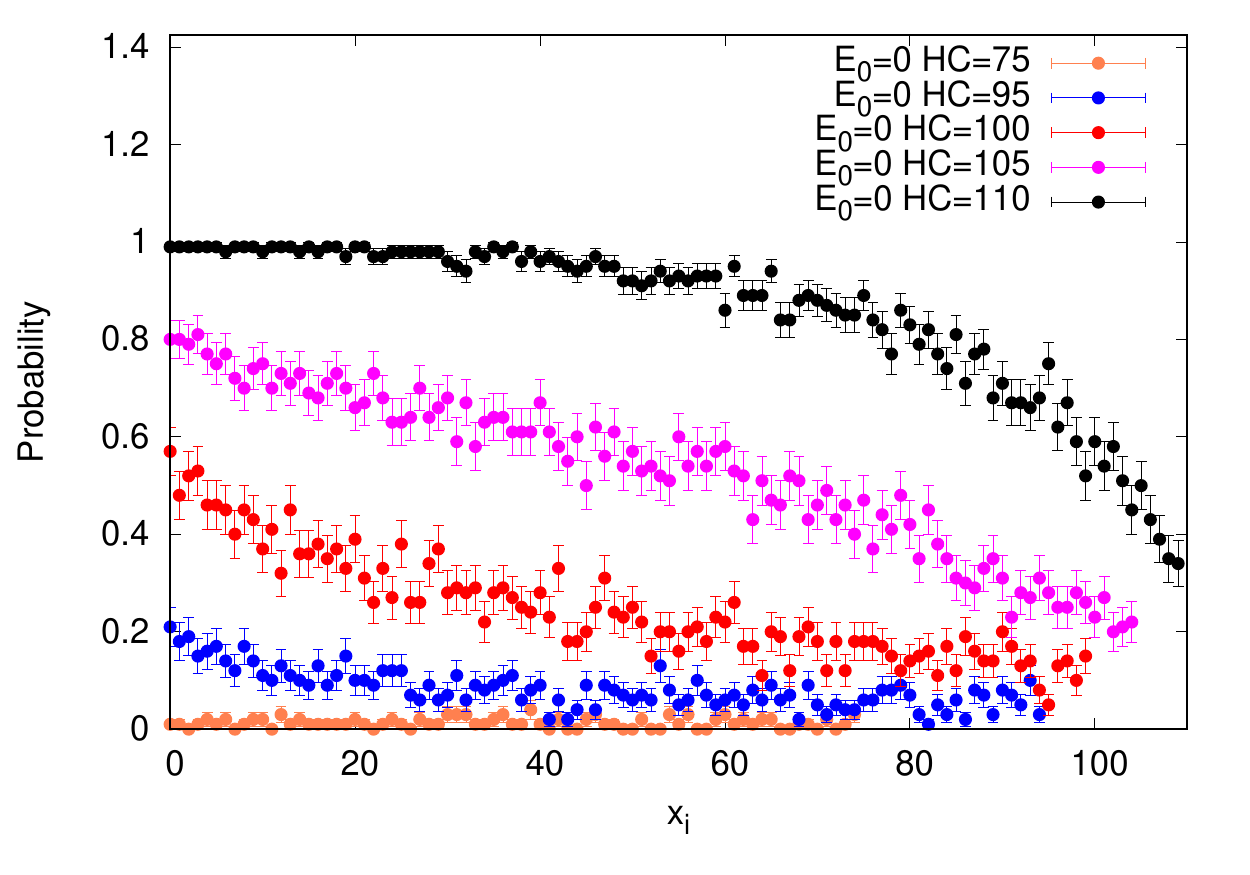}
\caption{The picture shows the probability that the leading components of the isolated eigenvector are contributed from hidden clique sites. Points are averages over $100$ graphs $G(N=10^4, p=0.5, K_{HC})$ with different values of $K_{HC}$, e.g. $K_{HC}=110$ black points, $K_{HC}=105$ magenta points,  $K_{HC}=100$ red points,  $K_{HC}=90$ blue points,  $K_{HC}=75$ coral points.}
\label{alonnohints}
\end{figure}

A completely connected graph of $n$ sites, $G(n,1.0)$, has an adjacency matrix of the same structure, with all $a_{ij} = 1$, and $0$ on the diagonal.  The eigenvalue for a uniform state lies at $\sqrt{n}$, but all the other eigenstates are degenerate in energy, with a negative eigenvalue close to zero.  To a first approximation, one can think of these eigenvectors as each consisting of $+1$ on a single site, and $-1/n$ on all the other sites of the subgraph, so that they will be orthogonal to the uniform eigenstate lying at a higher energy.  In the tight-binding language, the states of a hidden clique will hybridize with the random graph's energy band when the hidden clique subgraph is embedded in the larger random $G(N,0.5)$.  In the case Alon et al. \cite{alon1998finding} used for a simple analysis, $K_{HC} = 10 \sqrt N$, most of the hidden clique states merge into the random graph band with no visible change in its density of states, but the two quasi-uniform states survive outside the band.  An overall uniform state is still seen at $0.5 \sqrt N$, and the state which was uniform over just the hidden clique is found a small distance outside the energy band, with a gap between it and the uppermost state in the energy band.  If we define $\alpha = K_{HC}/  \sqrt N$, the gap disappears when $\alpha$ lies between $1.1$ and $1.2$, as shown in Fig.  \ref{gap}. The eigenstate just outside the gap has its largest contributions on the planted clique, with all these components of the same sign, and can thus be used to identify the clique sites. Once it joins the energy band, its dominant components come from both hidden clique and regular sites of the graph.

In Fig. \ref{distributioncomp}, we show the distribution of the eigenvector components on the hidden clique sites for a case with $\alpha = 1.5$  and compare this distribution with the distribution of the components of that eigenvector on the other sites of the graph.  They can clearly be distinguished.  
Until $\alpha = 1.2,$ the hidden clique sites are easily separated out, but at and below $1.1$, their magnitudes no longer clearly distinguish them.  
Fig. \ref{distributioncomp-100} (\textbf{a}) shows how this happens at $\alpha = 1.0$. 
If we sort the components of this eigenvector into a list, and observe the probability that a component comes from a hidden clique site, as a function of its order in the list, we see (in Fig. \ref{alonnohints}) that at $\alpha = 1.1$ and $1.05$, more than half of the larger components surely mark hidden clique sites, so that a subsequent cleanup stage should identify the hidden clique.  At $\alpha = 1$ and below, this becomes more difficult. At $\alpha = 1.1,$ about $90\%$ of the hidden clique sites are still found in the largest components of the highest eigenstate of the energy band, but at $\alpha = 1.0,$ this has dropped to $25\%$, and at $\alpha = 0.95,$ only $8\%$  survive.

 \begin{figure}[t]
\centering
\subfloat[]{\label{fig:mdleft}{\includegraphics[width=1\columnwidth, keepaspectratio=true, angle=0]{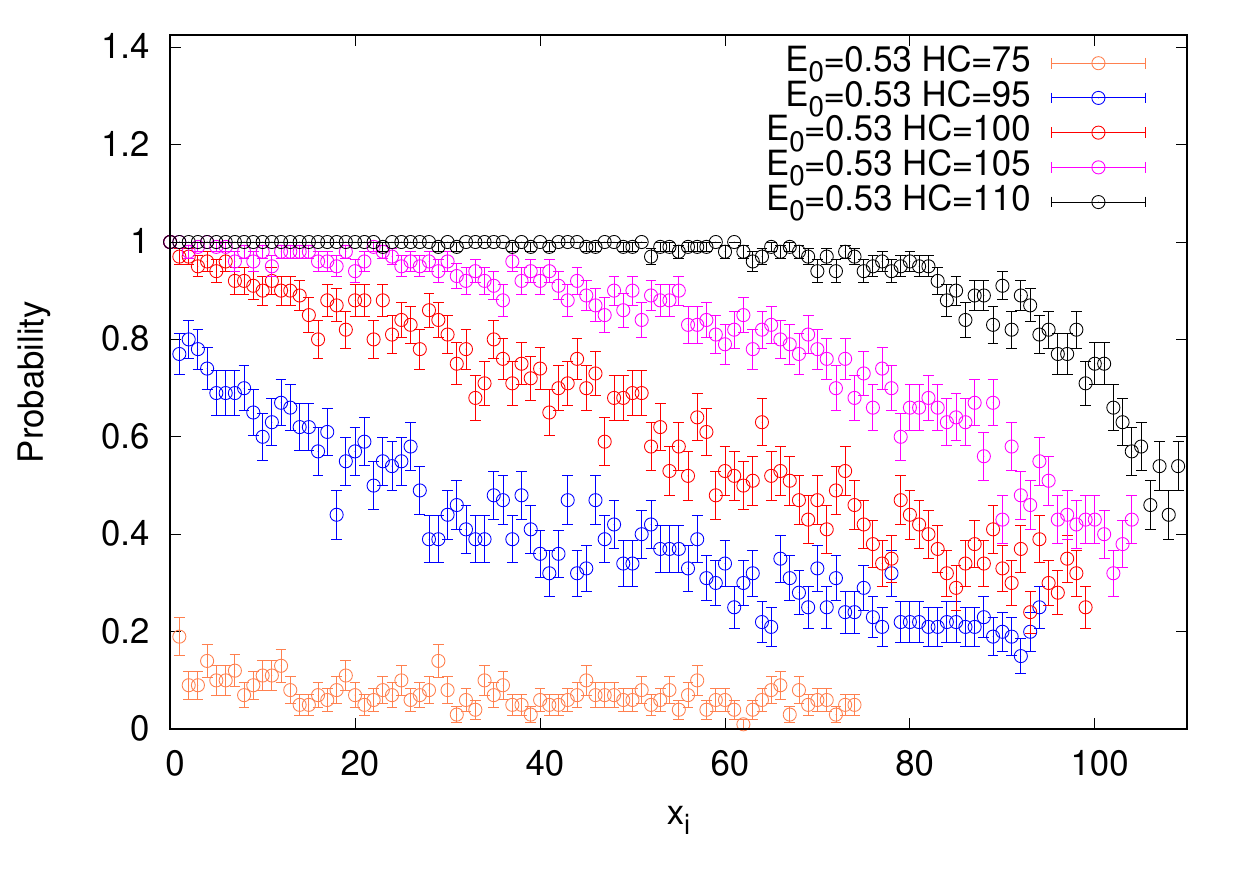}}}\hfill
\subfloat[]{\label{fig:mdright}{\includegraphics[width=1\columnwidth, keepaspectratio=true, angle=0]{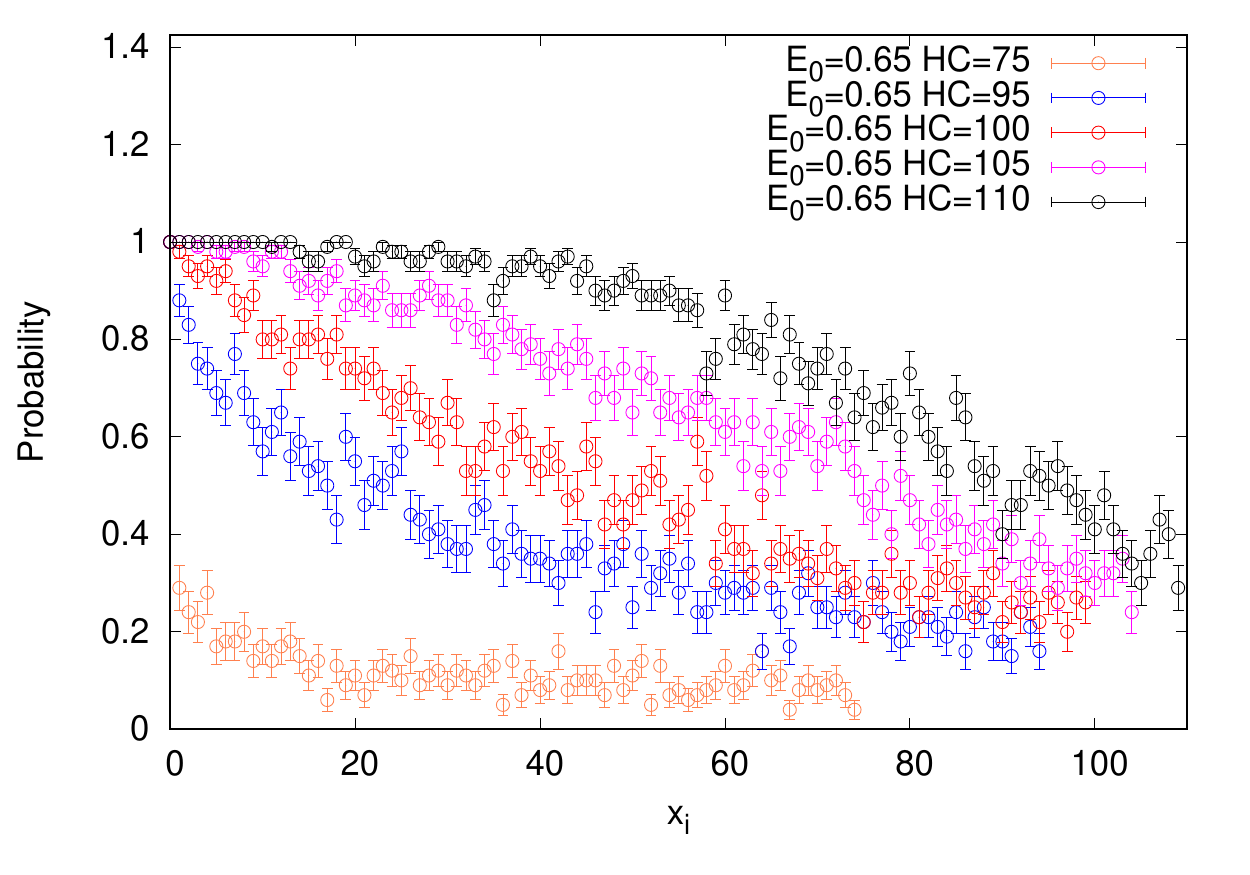}}}
\caption{(\textbf{a}) and  (\textbf{b})   show the probability that the leading components of the isolated eigenvector are contributed from hidden clique sites with $E_0=0.53$ and $E_0=0.65$ respectively. Points are averages over $100$ graphs $G(N=10^4, p=0.5, K_{HC})$ with different values of $K_{HC}$, e.g. $K_{HC}=110$ black empty points, $K_{HC}=105$ magenta empty points,  $K_{HC}=100$ red empty points,  $K_{HC}=90$ blue empty points,  $K_{HC}=75$ coral empty points.}
\label{figsalonhints}
\end{figure}

Exploiting the \textit{tight binding} interpretation of the energy levels of the planted clique/random graph mixture gives us a better way to study this regime. As a hint, we pick one site, site $0$, known to be in the planted clique, and raise its energy to move it towards or even beyond the upper edge of the band of states, by adding a constant term, $E_0$, to $a_{00}$.  This delays the disappearance of the gap between the isolated state and the upper edge of the band as we consider planted cliques of decreasing size.  The hope is that the \textit{impurity level} that remains outside the band for our inspection will combine a dominant component located on site $0$ with a cloud of sites from the hidden clique surrounding it.  

Fig. \ref{distributioncomp-100} (\textbf{b}), with $E_0 = 0.6$ and $\alpha = 1.0$, shows this effect.  Compare the degree of separation when the energy of site $0$ is boosted with the loss of separation shown in Fig. \ref{distributioncomp-100} (\textbf{a}).  An online GIF, which shows the evolution of the wave function coefficients as alpha decreases from $1.5$ down to $0.5$, and shows the two cases $E_0 = 0$ and $E_0 = 0.5$ side by side, makes this clear but could not be included in the published manuscript.  We make this available as supplementary material.

Values of $a_{00}$ ranging from $0.5$ to $0.7$ accomplish this, as Figs \ref{figsalonhints} (\textbf{a}) and (\textbf{b}) show.  Fig. \ref{figsalonhints} (\textbf{a})  repeats the analysis of Fig. \ref{distributioncomp} for five sizes of $HC$, using $E_0 = 0.53$, and makes $1.1$ to $1.0$ easy to extract.  Fig. \ref{figsalonhints}  (\textbf{b})  uses $E_0 =0.63$, and loses some of the sites that were seen before at $\alpha = 1.0 $ and above, but extracts more of the sites from $0.95$ and $0.75$.  For $\alpha = 1.0$, setting $E_0 = 0.53$ yields $63\%$ of the clique sites and at $\alpha = 0.95$ we obtain $39\%$ of the planted sites.  Increasing $E_0$ slightly to 0.65 decreases the number of sites that can be identified for $\alpha$ of 0.95 and above, but allows us to extract $13\%$ of the sites in the planted clique with $\alpha = 0.75$.  This seems to be the lower limit that we can uncover with this trick. 

Introducing a hint as we have done will reduce the number of sites which any local search must consider by roughly one half, since sites that are not linked to site $0$ can be ignored.  Perhaps the greater sensitivity that this trick gives to the spectral methods is simply the effective reduction of the limiting value of $\alpha$ for which they work by a factor of $1/\sqrt{2}.$  But the improvement in the ability of the method to identify planted clique sites for $\alpha$ above the higher limit is also a significant benefit.  
Increasing $E_0$ to still larger values creates an impurity level well outside the energy band, but without a cloud of sites surrounding it from the planted clique.  In that limit any site directly linked to site $0$ is equally likely to be seen with a large component in this eigenstate. 

\subsection{Iterative methods}

Next we consider methods of searching for the hidden clique that involve iteration. We shall employ two approaches, the $SM^1$ greedy algorithm with a simple modification, and the belief propagation scheme introduced by Deshpande and Montanari \cite{deshpande2015finding}.  First, we must make a further modification of the adjacency matrix.  We will use $\mathbf{\tilde{A}}$, whose elements are defined by: 

\begin{equation*}
\begin{split}
&\tilde{A}_{ij}  =  { \tilde{a}_{ij}  },\\
& \tilde{a}_{ij} = \tilde{a}_{ji} ,
\end{split}
\end{equation*}

where $\tilde{a}_{ij}=1$  if the link is present,  $\tilde{a}_{ij}=-1$  if the link is absent, and   $\tilde{a}_{ii}=0$.

The reason for the extra nonzero entries is simple.  It generates the same energy band, with double the width, and moves the special uniform state at $0.5 \sqrt{N}$ into the center of the band, where it no longer interferes with constructing eigenstates at the top of the energy band by iterative techniques such as the power method. The power method  \cite{mises1929praktische,ipsen2005analysis,lei2016coordinate} is an iterative algorithm that returns the greatest eigenvalue of a diagonalizable matrix $\mathbf{\tilde{A}}$ and the corresponding eigenvector  $\vec{s}$.   The power method applies the adjacency matrix $\mathbf{\tilde{A}}$ to a unit vector repeatedly until the result converges.  The result $\vec{s}$ is the eigenvector corresponding to the largest eigenvalue of $\mathbf{\tilde{A}}$, and this now becomes the special eigenvector which is dominated by the largest (planted) clique.  Although the power method allows analysis of larger graphs than we could study with Armadillo \cite{sanderson2016armadillo}, we did not find that it was any faster for our purposes.  The matrix $\mathbf{\tilde{A}}$, however, is the basis for the beliefs introduced in  \cite{deshpande2015finding}, and this proves capable of finding planted cliques still smaller than those exposed in Figs \ref{alonnohints} and \ref{figsalonhints}.


	To complete this analysis, we have also  considered the \textit{approximate message-passing} (\textbf{AMP}) algorithm given by Deshpande et al. in \cite{deshpande2015finding}. They developed a rigorous analysis that is asymptotically exact as $N \to \infty$ and they prove that their algorithm is able to find hidden cliques of size $K_{HC} \ge \sqrt{N/e}$ with high probability. 
	 \textbf{AMP} is derived as a form of \textit{belief propagation} (\textbf{BP}), a heuristic machine learning method for approximating posterior probabilities  in graphical models. \textbf{BP} is an algorithm \cite{montanari2007solving,felzenszwalb2006efficient,yedidia2001generalized,AngeliniTersenghi}, which extracts marginal probabilities for each variable node on a factor graph. It is exact on trees, but was found to be effective on loopy graphs as well \cite{deshpande2015finding,mezard2009information,frey1998revolution,mooij2005sufficient}. It is an iterative message passing algorithm that exchanges messages from the links to the nodes, and from them it computes marginal probabilities for each variable node.  When the marginal probability has been found, as \textbf{BP} has converged, one can obtain a solution of the problem, sorting the nodes by their predicted marginal probabilities. However it is possible, if the graph is not locally a tree, that \textbf{BP}  does not find a solution or converges to a random and uninformative fixed point. In these cases the algorithm fails.  \textbf{BP} for graphical models runs on factor graphs where each variable node is a site of the  original graph $G(N,p)$, while each function node is on a link of the original graph $G(N,p)$.
	Here we describe briefly the main steps that we have followed in implementing \textbf{AMP} algorithm.  For details we refer the reader to \cite{deshpande2015finding}.
	\textbf{AMP} runs on a complete graph described by an adjacency matrix  $\mathbf{\tilde{A}}$. \textbf{AMP}  iteratively exchanges messages from links to nodes, and from them it computes quantities for each node. These quantities represent the property that a variable node is, or not, in the planted set. It is intermediate in complexity and compute cost between local algorithms, such as our greedy search schemes, and global algorithms such as the spectral methods of Alon et al. \cite{alon1998finding}
For our purpose, we implemented a simple version of the algorithm in \cite{deshpande2015finding}, using Deshpande et al \cite{deshpande2015finding} equations. Here, we recall them:
	
	\begin{equation}
	\label{messages}
	\begin{split}
	&\Gamma^{t+1}_{i \to j}=\log \frac{K_{HC}}{\sqrt{N}}+\sum^{N}_{l\neq i,j}\log\left( 1+\frac{(1+\tilde{A}_{l,i}) \text{e}^{\Gamma^{t}_{l\to i}}}{\sqrt{N}} \right) +\\
	&-\log\left( 1+\frac{\text{e}^{\Gamma^{t}_{l\to i}}}{\sqrt{N}}\right),
	\end{split}
	\end{equation}
	
	\begin{equation}
	\label{marginal}
	\begin{split}
	&\Gamma^{t+1}_{i}=\log \frac{K_{HC}}{\sqrt{N}}+\sum^{N}_{l\neq i}\log\left( 1+\frac{(1+\tilde{A}_{l,i}) \text{e}^{\Gamma^{t}_{l\to i}}}{\sqrt{N}} \right)+\\
	&-\log\left( 1+\frac{\text{e}^{\Gamma^{t}_{l\to i}}}{\sqrt{N}}\right).
	\end{split}
	\end{equation}
	
	Equations (\ref{messages}) and (\ref{marginal}) describe the state evolution of messages and vertex quantities $\Gamma^{t}_{i}$. They run on a fully connected graph, since both the presence or absence of a link between sites is described in the adjacency matrix $\mathbf{\tilde{A}}$. For numerical stability, they are written using logarithms. Initial conditions for messages in  (\ref{messages}) are randomly distributed and less than $0$.
	 The constant part is obtained by observing that relevant scaling for hidden clique problems is $\sqrt{N}$.  
	
	Equation (\ref{messages}) describes the numerical updating of the outgoing message from site $i$ to site $j$. It is computed from all ingoing messages to $i$, obtained at previous iteration, excluding the  outgoing message from $j$ to $i$. These messages, i.e. equation (\ref{messages}),  are all in $ \mathbb{R}$ and they correspond to so-called odds ratios that vertex $i$ will be in the hidden set $C$. In other words, the message from  $i$ to $j$ informs  site $j$ if site $i$ belongs to the hidden set or not, computing the odds ratios of all remaining  $N-2$ $l$ sites of the graph, with $l\neq i,j$. When a site $l$ is connected to site $i$, the difference between logarithms, in the sum,  will be positive and will correspond to the event that the site $l$ is more likely to be a site of $C$ than a site outside it.  However, when $l$ is not connected to $i$ the corresponding odds ratios will be less than one, i.e. the difference of logarithms, in the sum of equation (\ref{messages}),  will be less than zero, and will correspond to the event that the site $l$ is more likely to be outside the hidden set. The sum of all the odds ratios will update equation (\ref{messages}),  telling us  if site $i$ will be more likely to be in $C$ or not.
	
	Equation (\ref{marginal}), instead, describes the numerical updating of the vertex quantity  $\Gamma^{t}_{i}$. It is computed from all ingoing messages in $i$, and  is an estimation of the likelihood that $i\in C$. These quantities are larger for vertices that are more likely to belong to the hidden clique \cite{deshpande2015finding}. Elements of the hidden set, therefore, will have $\Gamma^{t_c}_{i\in C}>0$, while  elements that are not in the hidden set will  have  $\Gamma^{t_c}_{i\not \in C}<0$.

	As iterative \textbf{BP}  equations,  (\ref{messages}) and (\ref{marginal})  are useful only if they converge. The computational complexity of each iteration is $\mathcal{O}(N^2)$, indeed, equation (\ref{messages}) can be computed efficiently using the following observation:
	
	\begin{equation}
	\label{trick}
	\begin{split}
	&\Gamma^{t+1}_{i \to j}=\Gamma^{t+1}_{i}-\log\left( 1+\frac{(1+\tilde{A}_{j,i}) \text{e}^{\Gamma^{t}_{j\to i}}}{\sqrt{N}} \right)+\\
	&+\log\left( 1+\frac{\text{e}^{\Gamma^{t}_{j\to i}}}{\sqrt{N}}\right).
	\end{split}
	\end{equation}
	
	The number of iterations needed for convergence for all messages/vertex quantities is of order $\mathcal{O}(\log N)$, which means that the total computational complexity of the algorithm is $\mathcal{O}(N^2 \log N)$. Once all messages in (\ref{messages}) converge, the vertex quantities given by (\ref{marginal}) are sorted into descending order. Then, the first $K_{HC}$ components are chosen and checked to see if they are a solution. If a solution is found we stop with a successful assignment, else the algorithm returns a failure. For completeness, our version of \textbf{AMP} algorithm  returns a failure also when it does not converge after $t_{\text{max}}=100$ iterations.

	As a first experiment we run simulations which reproduce the analysis in \cite{deshpande2015finding}, but apply their methods to a larger sample, $N = 10^4$.  In Fig. \ref{AMPVSSM1} we show the results of fraction of successful recovery by \textbf{AMP} after one convergence, as a function of $\alpha$. As the analysis in \cite{deshpande2015finding} predicts, the \textbf{AMP} messages converge down to about $\alpha = \sqrt{N/e}$, but with a decreasing probability of convergence, or with success in a decreasing fraction of the graphs that we have created.  At and below the algorithmic threshold of \textbf{AMP} for this problem, we obtained very few solutions.

\subsection{Greedy search with early stopping}

\begin{figure}
\includegraphics[width=1\columnwidth, keepaspectratio=true, angle=0]{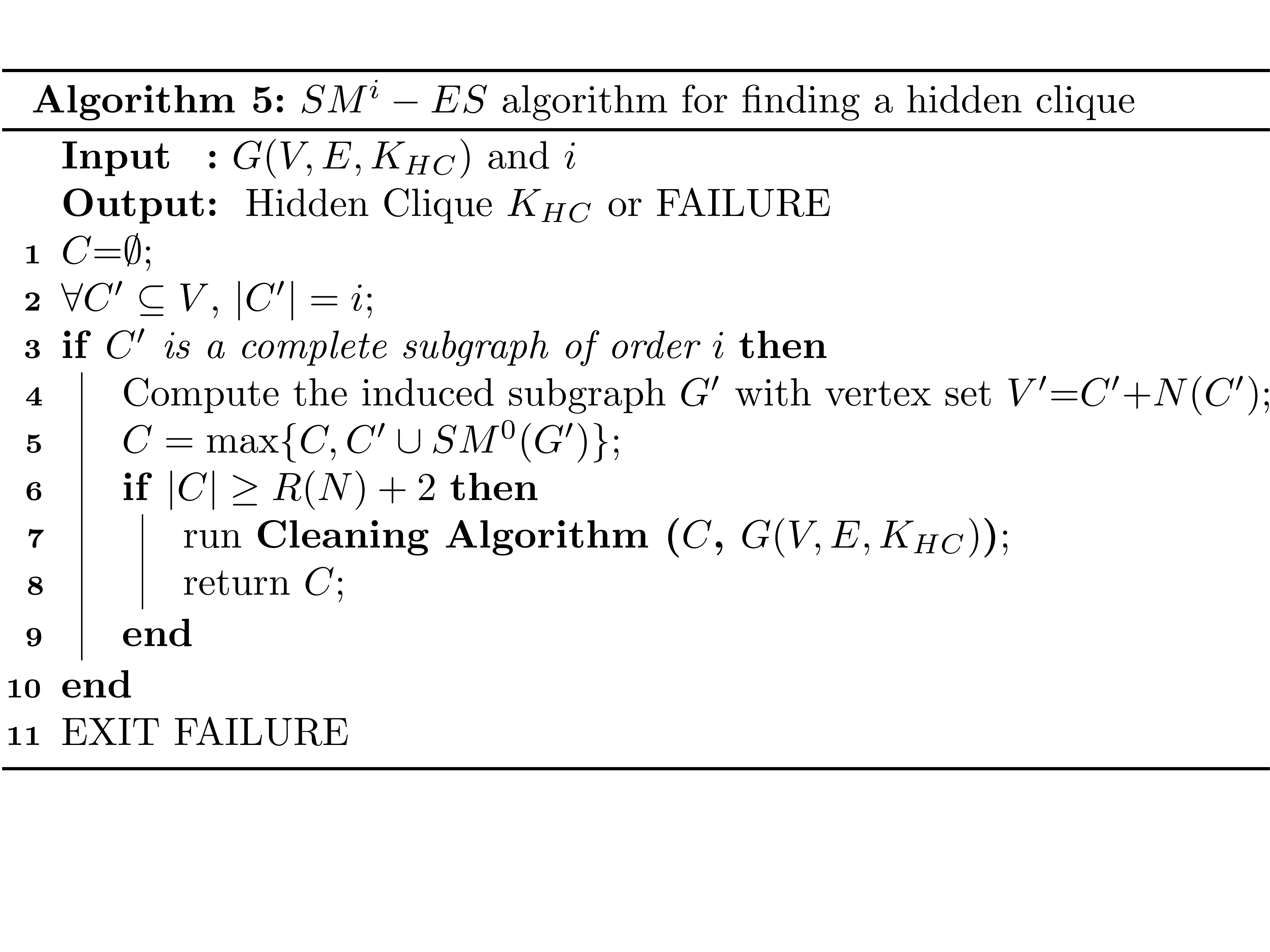}
\label{algoSM1ES}
\end{figure}
\begin{figure}
\includegraphics[width=1\columnwidth, keepaspectratio=true, angle=0]{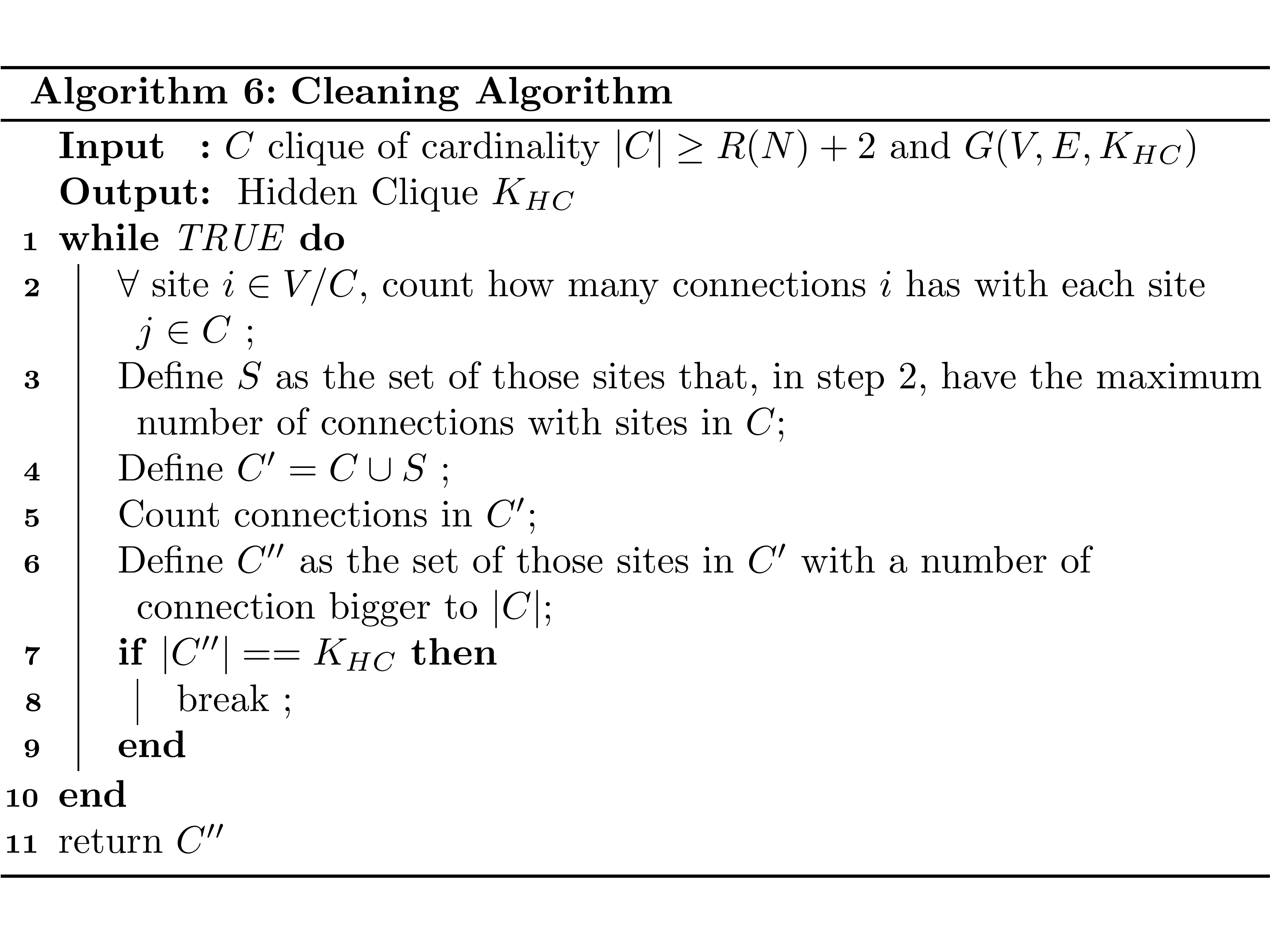}
\label{cleaning}
\end{figure}

\begin{figure}
\centering
\includegraphics[width=1\columnwidth, keepaspectratio=true, angle=0]{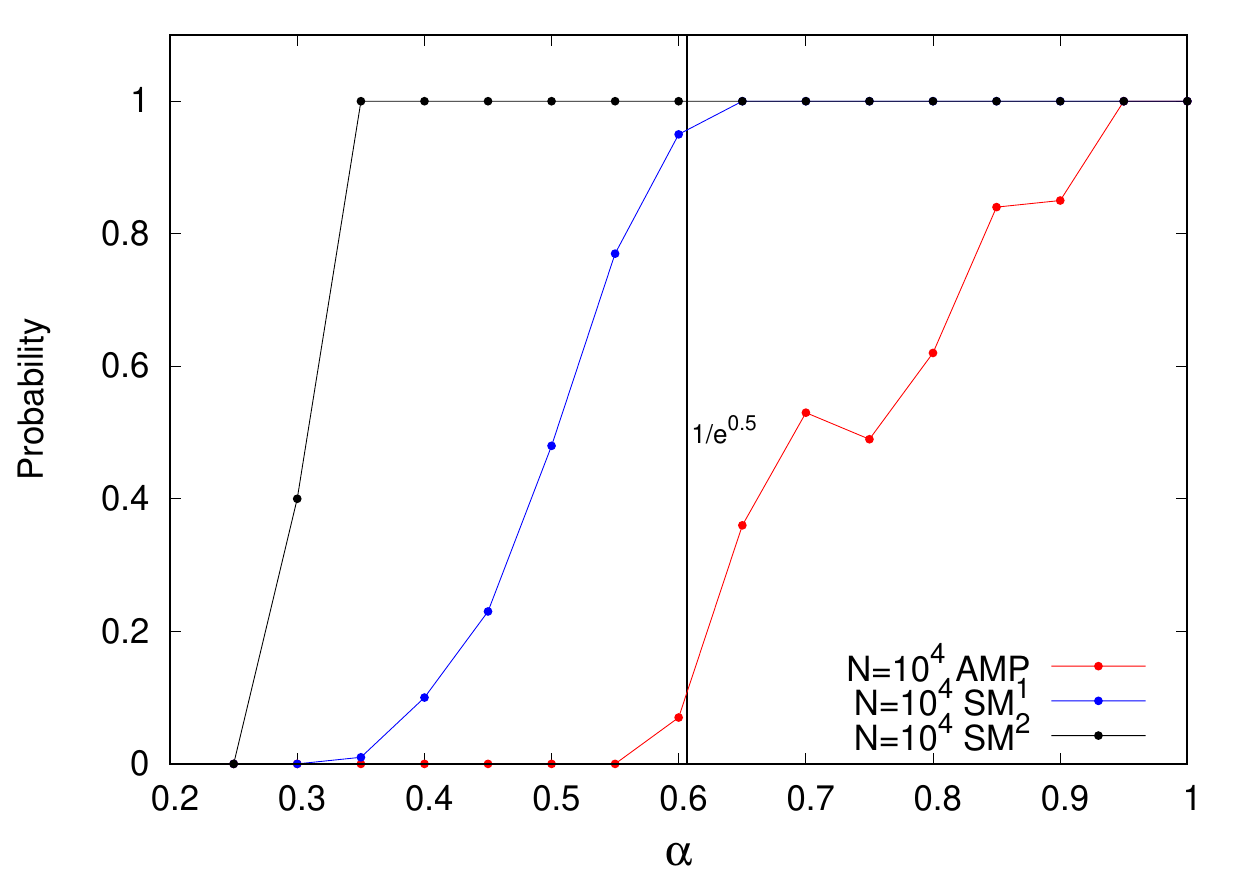}
\caption{The picture shows the probability of success for recovering the hidden clique as function of $\alpha$ for \textbf{AMP} (red points), $SM^1$ with early stopping (blue points) and $SM^2$ with early stopping (black points). 
Points are averages over $100$ graphs $G(N=10^4, p=0.5, K_{HC}=\alpha \sqrt{N})$ for \textbf{AMP} algorithm and $SM^1$ with early stopping, while they are an average over $5$ graphs for $SM^2$ with early stopping. }
\label{AMPVSSM1}
\end{figure}

We also explored using our greedy search methods to uncover a planted clique in this difficult regime.  Our hypothesis was that using $SM^0$ was unlikely to succeed since almost all sites selected at random do not lie within the planted clique.  But $SM^1$ seems more promising, even with its $\mathcal{O}(N^3)$ cost.  And if the search gave rise to any clique of size $R(N)$ or larger, perhaps by a fixed amount $d \ge 2$,  that is strong evidence of the existence of the planted clique.  A clique of this size is a reliable starting point for a cleanup operation to find the remaining sites, using Algorithm $6$.  To our surprise, as shown in Fig. \ref{AMPVSSM1}, this succeeds in a greater fraction of the graphs than does \textbf{AMP} for planted cliques, when $\alpha < 1$.  This strategy of stopping $SM^1$ as soon as the hidden clique is sufficiently exposed to finish the job with the clean up Algorithm $6$ produces the hidden clique almost without exception in our graphs of order $10^4$ through the entire regime from $\alpha = 1$  down to $\alpha = 1/\sqrt{e}$.  In this regime, \textbf{AMP},  converges to a solution in a rapidly decreasing fraction of the graphs.  We studied the same $100$ graphs with $\mathbf{AMP}$ as were solved with $SM^1$ at each value of $\alpha$.  Using $SM^1$ with early stopping, we could extract planted cliques as small as $\alpha = 0.4$.

\begin{figure}
\centering
\includegraphics[width=1\columnwidth, keepaspectratio=true, angle=0]{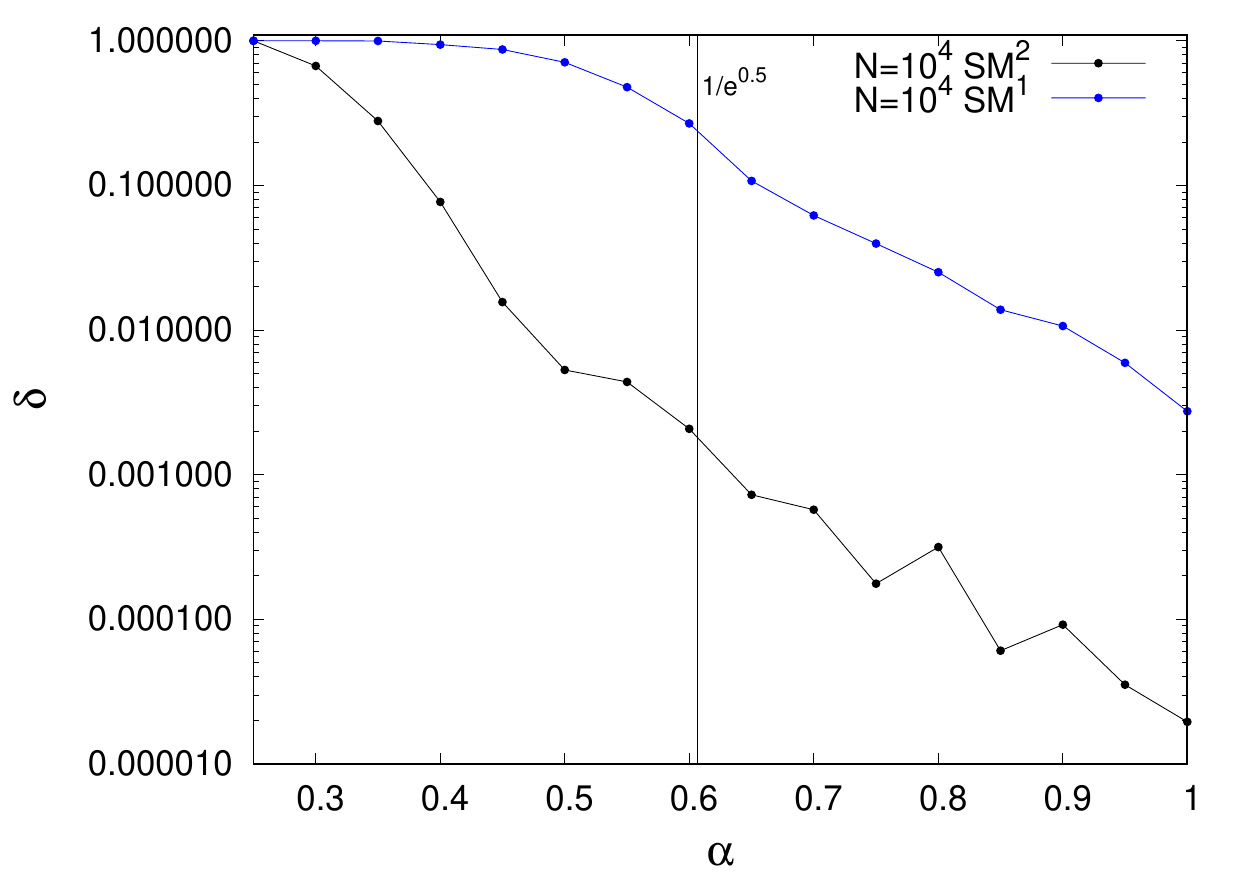}
\caption{The picture shows the fraction of iteration ($\delta$) that $SM^1$ with early stopping (blue points) and $SM^2$ with early stopping (black points) need to make before stopping, in  $\log$ scale,  as function of $\alpha$. Points are averaged over $100$ and $5$ graphs $G(N=10^4, p=0.5, K_{HC}=\alpha \sqrt{N})$, for $SM^1$ and $SM^2$, respectively, both with early stopping.}
\label{NS-SM1}
\end{figure}

\begin{figure}
\centering
\includegraphics[width=1\columnwidth, keepaspectratio=true, angle=0]{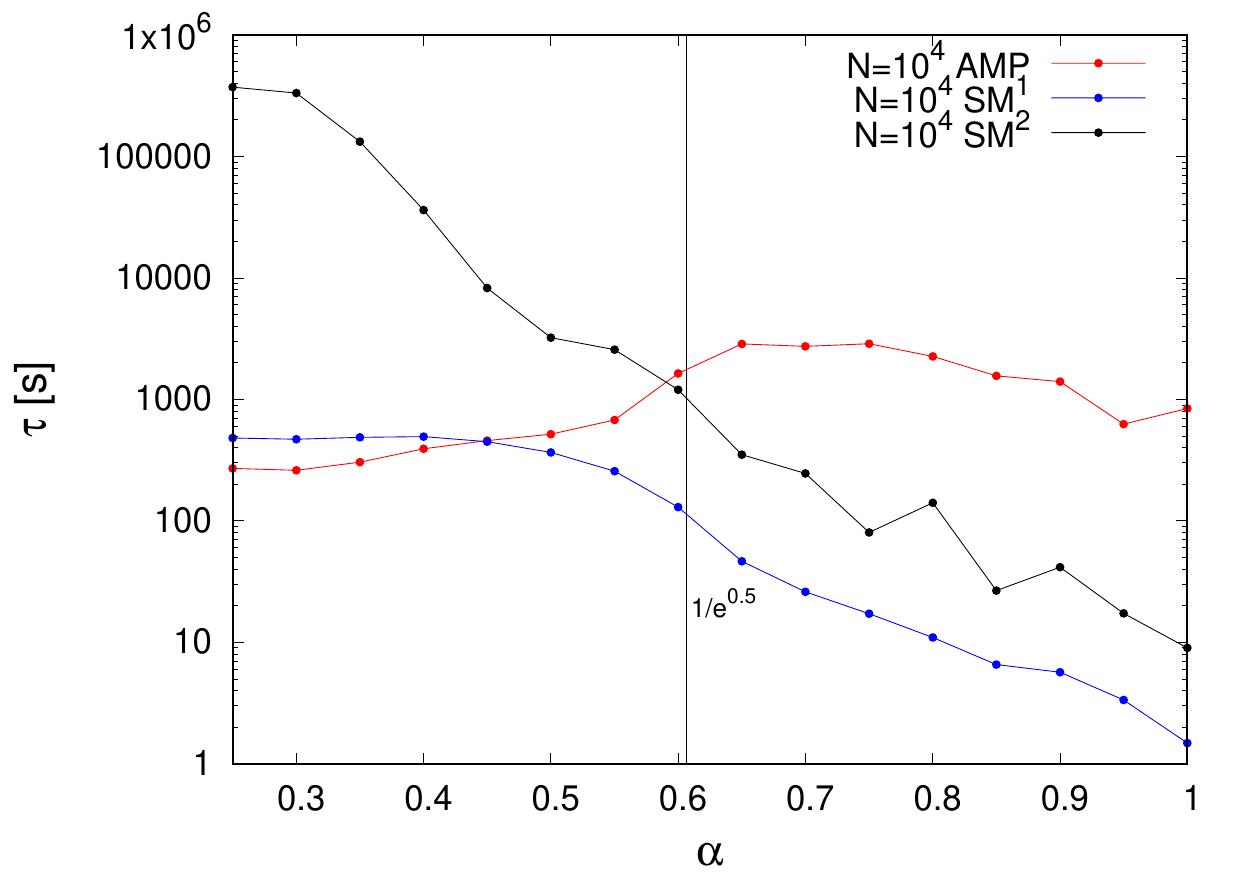}
\caption{The picture shows as function of $\alpha$ the average time, in seconds, needed by \textbf{AMP} (red points), and $SM^1$ (blue points) and  $SM^2$ (black points) with early stopping, to give an answer on a single graph $G(N=10^4, p=0.5, K_{HC}=\alpha \sqrt{N})$.}
\label{TIME}  
\end{figure}

Naturally, the success of early stopping in making $SM^1$ useful led us to try the same with $SM^2$.  We tried this with only $5$ graphs at each value of $\alpha$, and were able to identify the planted clique in all graphs down to $\alpha = 0.35$, and in two out of five graphs at $\alpha = 0.3$.  The method was not successful at all at $\alpha = 0.25$.  The third curve of results in Fig. \ref{AMPVSSM1} shows the results of the three methods.  It appears that the \textit{local}, greedy methods, when used repeatedly in this fashion, are actually stronger than the more globally extended survey data collected by \textbf{AMP}.  But to compare their effectiveness, it is also necessary to compare their computational costs. This is explored in Figs. \ref{NS-SM1} and \ref{TIME}.


In Figs \ref{NS-SM1} and \ref{TIME}, we compare the effectiveness of $SM^1$ and $SM^2$ with early stopping and \textbf{AMP}.  First, we find that the number of trials required for $SM^1$ to expose the hidden clique was close to $N$ at the lowest successful searches, but dropped rapidly (the scale is logarithmic) for $\alpha > \sqrt{1/e}$.  As $\alpha$ approaches $1$, there are more starting points than there are points in the hidden clique, while for $\alpha < \sqrt{1/e}$, not every point in the hidden clique is an effective starting point. The plot in Fig. \ref{NS-SM1} expresses the number of sites which needed to be searched as $\delta$, the fraction of the search space traversed.  Similarly, we plot as $\delta$ for $SM^2$ the number of links searched divided by the number of links in the graph.  

We briefly explored the importance of where to stop the search by running $SM^1$ to completion for a small number of graphs at $\alpha = 0.6$ and considering the sizes of the cliques found.  This distribution varies quite widely from one graph to another.  The full hidden clique is frequently found, and the most common sized results were about half of the hidden clique size.  Only a very few cliques returned by $SM^1$ were within $1-4$ sites of $R(N)$, so we recommend the stopping criterion $R(N) + 2$ as a robust value.

The average running time to solve one graph for each of the three is plotted in Fig. \ref{TIME}. The average cost of solving \textbf{AMP}, (red points) is greatest just above $\alpha = 1/\sqrt{e}$ where it sometimes fails to converge, and decreases at higher $\alpha$, largely because convergence is achieved, with fewer iterations as $\alpha$ increases.  The cost decreases at lower values of $\alpha$ because \textbf{AMP} converges more quickly, but this time to an uninformative fixed point.   $SM^1$ with early stopping (blue points) requires less time than \textbf{AMP} to expose the planted clique at all values of $\alpha$ where one or both of the methods are able to succeed, and is several hundred times faster at $\alpha = 1$.   $SM^2$ with early stopping (black points) is more expensive than  $SM^1$ with early stopping at all values of $\alpha$, but is also less costly than \textbf{AMP} in the range $1/\sqrt{e} < \alpha <1$.  It is the most costly algorithm at still lower values of $\alpha$, but the only method that can provide any solutions down to the present lower limit of $\alpha = 0.3$.  This efficiency, as well as the ability of local greedy algorithms with early stopping to identify cliques with $\alpha < \sqrt{1/e}$, is a surprising and novel result.

	
\section{Conclusions}
\label{sec:Conclusions}

More than $20$ years have elapsed since the DIMACS community reviewed algorithms for finding maximum cliques (and independent sets) in Erd{\"o}s-R{\'e}nyi graphs $G(N,p)$ with $N$ sites and bonds present with fixed probability, $p$.  Computer power and computer memory roughly $100\times$ what was available to the researchers of that period are now found in common laptops. But unfortunately, the size of the problems that this can solve (in this area) only increases as the $\log$ of the CPU speed.
We can now explore the limits of polynomial algorithms up to $N=10^5$, while the DIMACS studies reached only a few thousand sites.  In contrast to problems like random $3$-SAT, for which almost all instances have solutions by directed search \cite{selman1993local} or belief propagation-like \cite{mezard2002analytic,mezard2007constraint,marino2016backtracking} methods which approach the limits of satisfiability to within a percent or less, finding a maximum clique remains hard over a large region of parameters for almost all random graphs, if we seek solutions more than a few steps beyond the dynamical threshold, $\log_2 N$ set by the naive greedy algorithm.  Using tests more detailed than the \textit{bakeoff}  with which algorithms have been compared, we show that expensive $\mathcal{O}(N^3)$ and $\mathcal{O}(N^4)$ searches can accurately reproduce the distribution of maximum clique sizes known to exist in fairly large random graphs.  (Up to at least $N = 500$ for the $\mathcal{O}(N^3)$ algorithm and about $N = 1500$ for the $\mathcal{O}(N^4)$ algorithm.) This is a more demanding and informative test of the algorithms' performance than seeing what size clique they each can extract from graphs whose actual maximum clique size is unknown.

A more promising approach is to use the simplest search algorithm to define a subgraph much smaller than $N$ as a starting subset in which to apply the higher order search strategies. This cannot produce the exact maximum clique, or even get within a percent or less of the answer as with SAT, because the naive initial search combines sites which belong in different maximum cliques into the starting set and the higher order follow-up search that we employ does not fully separate them.  Nonetheless, extrapolating our several algorithms towards the scales that future data and future computing power will afford suggests that the challenge of exceeding the dynamical threshold can be met for $N$ at least $10^{10}$ and perhaps up to values such as $10^{20}$.  These are in the range presented by the information retrieval challenges of modern commercial data.

The second challenge we considered is locating and reconstructing a hidden clique, perhaps with the use of a hint.  The hint (a site known to be in the hidden clique) has the effect of eliminating parts of the graph that will not be in the full clique to the point that the eigenvector of the adjacency matrix corresponding to its largest eigenvalue can be used to identify the remaining clique members, or that greedy search in the remaining graph can extract a large fraction the hidden clique, with a  cleanup step used to identify the rest. Using spectral methods (augmented with a hint), Deshpande and Montanari's \cite{deshpande2015finding} \textbf{AMP}, or our slightly more than linear cost polynomial $SM^1$ with early stopping, we can reconstruct the hidden clique well within the text{challenge} regime that Jerrum pointed out.  What is surprising is that a version of the  the simplest greedy algorithm performs even better on problems of the largest currently achievable sizes.

The challenges posed at the start of this paper apply only as $N \to \infty$, in a problem with significant and interesting finite-size corrections.  Although computing power, data storage, and the data from which information retrieval tools are sought to find tightly connected communities all increase at a dramatic pace, all of these presently lie in the  \textit{finite-sized} range of interest, not at the asymptotic limit. Yet they are well beyond the scale of previous efforts to assess algorithms for this problem.  Since asymptotic behavior is  only approached logarithmically in the clique problem, we think that additional challenges of value should be posed in the finite size regime.  We have shown that effective searches for cliques can be conducted on graphs of up to $10^5$ sites, using serial programs.  With better, perhaps parallel algorithms, and the use of less-local search strategies such as \textbf{AMP}, can this sort of search deal with information structures of up to $10^9$ nodes using today's computers?  With computational resources of the next decade, and perhaps a better understanding of the nature of search in problems with such low signal-to-noise ratios as MaxClique, can we hope to see graphs of order $10^{12}$ being handled?

The criterion that we used to evaluate the $SM^{i}$ family of algorithms and their derivatives can be applied at the steps for any larger $N$, where $K_{\text{max}}$ increases by one.  For a graph in $G(N,0.5)$, constructed at the step rise, find the half of the  graphs which contain a clique of the size characteristic of the upper step.  Or show that the probability of seeing any graph with a clique of this size is greater than $\epsilon$. 

In finding hidden cliques in commercial data, use of some hints is reasonable, since communities in social data are defined by known exemplars.  For cliques of size $\mathcal{O}(\sqrt{N})$, can the hidden clique be restored with only one hint when it's size is $\epsilon \sqrt{N}$ for arbitrarily small $\epsilon>0$ at some affordable cost $c(\epsilon)$.  Can a constant number of hints or perhaps a fraction of $\log_2 N$ hints be used to reveal a planted clique whose size is only a small multiple of $K_{\text{max}}$, without reducing the cardinality of the sites in the graph to be searched to the square of the number of clique sites still to be found? i.e., without making the reconstruction search trivial by making its "$\alpha$" $<<1$?  A promising parallel approach to such \textit{needle in haystack} searches could be constructed by conducting many local searches, pruning their cost by early stopping, and then expanding on the most successful, as we have done in both searching for naturally occurring cliques and for planted solutions.

\section{Acknowledgements}
We enjoyed stimulating conversations with Federico Ricci-Tersenghi and Maria Chiara Angelini at the outset of this work.  RM and SK are supported by the Federman  Cyber Security Center of the Hebrew University of Jerusalem.

\bibliographystyle{ieeetr}
\bibliography{Bib}

\end{document}